%% file: main.tex
\documentclass{aimsessay}
\usepackage[utf8]{inputenc}
\usepackage[round]{natbib}
\usepackage{glossaries}
\makeglossaries
\input{glossary.tex}
\usepackage{arabtex}
\usepackage{utf8}
\setcode{utf8}
\usepackage{booktabs} 
\usepackage{array}    
\usepackage[below]{placeins} 
\usepackage{float}  
\usepackage{caption} 
\usepackage{subcaption} 
\usepackage{moreverb}   
\usepackage{hyperref}
\usepackage{epstopdf}
\usepackage{psfrag}
\usepackage{graphics}
\swapnumbers 
\theoremstyle{plain}

\theoremstyle{definition}

\numberwithin{equation}{section}
\title{\includegraphics[scale=0.7]{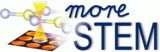}\hspace*{7.5cm}\includegraphics[scale=0.2]{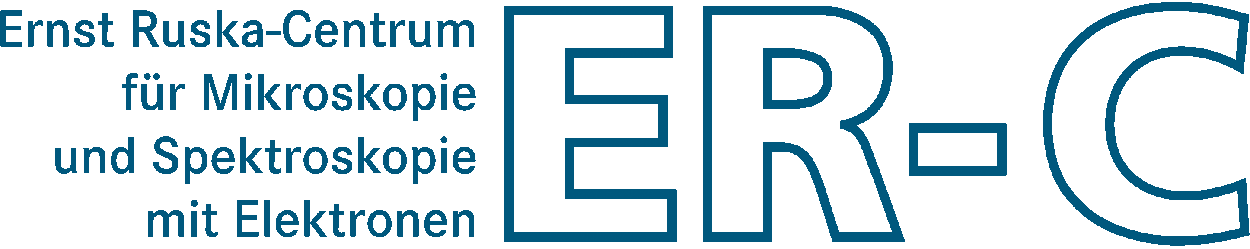}\\\vspace*{0.5cm}
Ptychographic Algorithms for Phase Recovery in 4D Scanning Transmission Electron Microscopy  }
\author{BY\\ [1cm]
AMEL SHAMSELDEEN ALI ALHASSAN  (aalhassa@ictp.it)\\
 Abdus-Salam International Centre for Theoretical Physics (ICTP) - Italy\\
 {\small Supervised by: Professor Doctor. Knut M\"{u}ller-Caspary}\\
 {\small RWTH University Aachen}\\
 {\small Mr. Hoel L. Robert }\\
 {\small RWTH University Aachen}\\
 {\small Doctor Tevfik Onur Mentes}\\
 {\small Elettra Synchrotron, Italy}\\
 }
\date{{\small August 2020}\\ [0.5cm]
   {\scriptsize\it AN ESSAY PRESENTED TO ABDUS-SALAM INTERNATIONAL CENTRE FOR THEORETICAL PHYICS IN PARTIAL FULFILLMENT OF THE REQUIREMENTS FOR THE AWARD OF POSTGRADUATE DIPLOMA IN CONDENSED MATTER}\\%
  \vspace{3cm}{\includegraphics[scale=1]{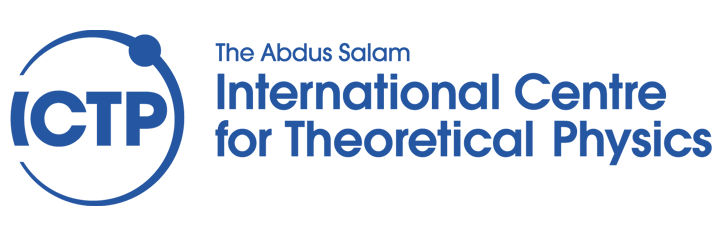}}}
\begin{document}
\pagestyle{empty}
\maketitle
%
\pagenumbering{roman}
\input{abstract}

\tableofcontents
\newpage
%
\pagenumbering{arabic}
\pagestyle{myheadings}
\input{STEM_anatomy}
\input{chapter4}
\input{chapter1new} 
\input{chapter3}
\input{results}
\printglossary[type=main]
\renewcommand{\bibname}{References}
\nocite{*}
\bibliographystyle{plainnat}
\bibliography{references}
\addcontentsline{toc}{chapter}{References}
\end{document}

%% file: glossary.tex
\newglossaryentry{resolution}
{
    name=resolution,
    description={Is the minimum distance between two points in the object than can be seen distinctly in the image. }
}

\newglossaryentry{specimen}
{
    name=specimen,
    description={The lattice or amorphous whose image is to be made with the microscope. It might be called "Sample" in other contexts.}
}

\newglossaryentry{point spread function}
{
    name=point spread function,
    description={ The Fourier transform of the exponential of the aberration function. $\mathcal{F}^{-1}[e^{i\chi (\vec{k})}](\vec{r}) \equiv h_0 (\vec{r})$ }
}

\newglossaryentry{diffractogram}
{
    name=diffractogram,
    description={Fourier transform of conventional TEM image }
}

%% file: abstract.tex
\abstract

In Momentum-resolved Scanning Transmission Electron Microscopy (4D STEM),
a convergent electron beam is raster-scanned across a think specimen in 2D
in real space. The corresponding 2D diffraction pattern, in momentum
space, to each point is recorded, forming a 4D data set. Information
decoding process can follow thereafter to produce an image of the specimen
in real space.

Ptychography is reconstruction algorithm that allow the extraction of the
probe wavefunction and the multiplicative object transmission function of
the specimen. Ptychography is implemented through direct and iterative
schemes. Some of which are the extended Ptychographic Iterative Engine
(ePIE), the Wigner Distribution Deconvolution (WDD) and the simpler
version of WDD, the Single Side-Band (SSB).

This thesis gives an overview of STEM ptychography giving examples of its
experimental and simulated implementations. The different ptychographic
reconstruction methods are explored in a mathematical framework when
applicable. Finally, an SSB reconstruction was made using an original
script for simulated data of MoS2 monolayer. Moreover, four-dimensional
data was recorded using a STEM instrument. A natural step following this
research would be the implementation of the WDD algorithm.

%% file: STEM_anatomy.tex
\begin{figure}
\hspace*{-1cm}
\centering
\includegraphics[scale=0.17]{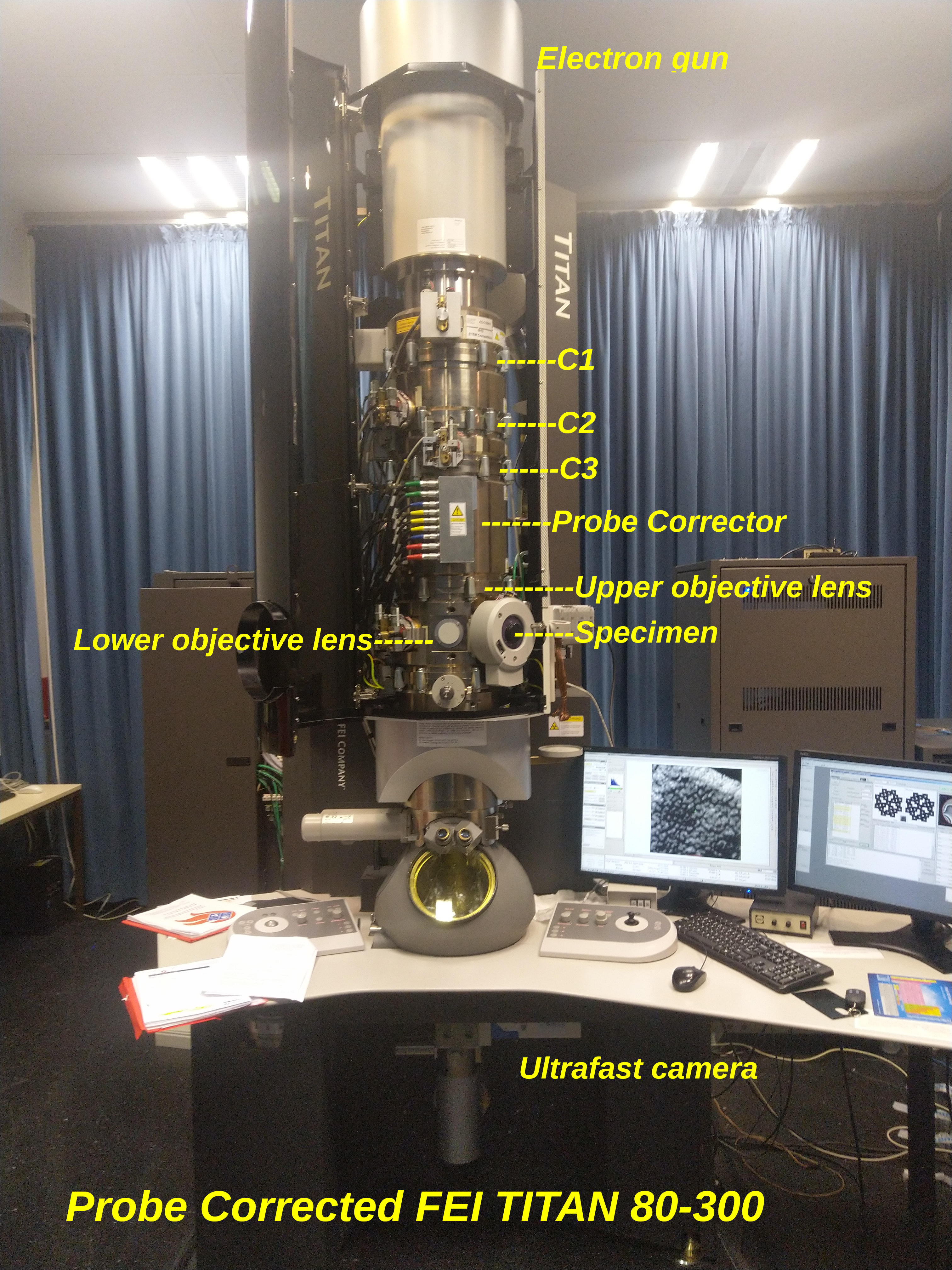}
\caption{STEM anatomy}
\end{figure}

%% file: chapter4.tex
\chapter{Introduction}

    \section{ conventional STEM imaging modes}

    Convergent illumination is used for STEM. Phase contrast results from the interference between scattered and unscattered beams.  It can be observed using an axial detector if the scattering angle is less than the radius of the incident cone. However, this yields a reduction in the overall phase contrast in the bright field imaging mode
    \citep{stemb}.
    Meanwhile, excluding the unscattered electrons, at $\theta > 100 mrad$, the resulting diffraction will be less sensitive to the crystalline nature of the specimen and more sensitive to the atomic number of its components. \citep{stemb}
    
    \begin{figure}[H]
    \centering
    \includegraphics[scale=0.3]{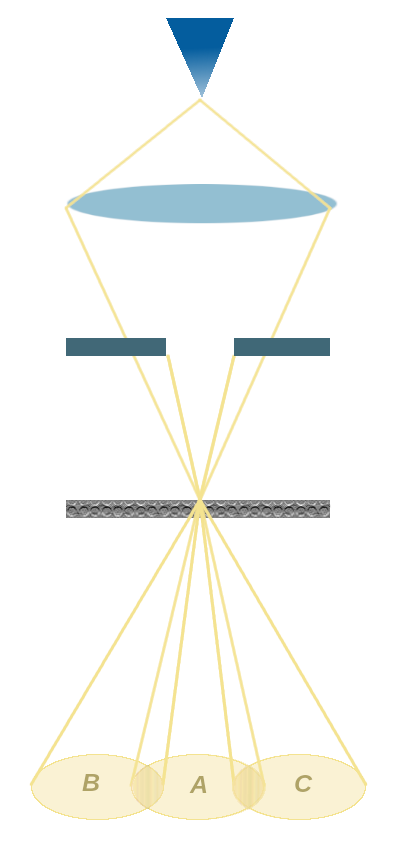}
    \caption{The overlapping between the transmitted beam A and the scattered beams B and C contains information about the relative phase between them. Moreover, the interference stripes in these areas has the same periodicity of the specimen. Raster-scanning the probe over the specimen allows the retrieval of the unique phase of diffraction pattern.   }
    \end{figure}

    \section{Established imaging modes in STEM}
    
    Transmission electron microscopes (TEM)
    are highly specialized for atomic resolution. Therefore, They are used intensely in material science. They can work on different modes, of which he two main modes are:
        
        \begin{itemize}
            
             \item  Bright-Field Imaging
            
                In STEM,the detector is disc-shaped. It is placed on the optical axis. It collects the diffraction pattern in the region of of the unscattered beam.
                
                If the specimen is very thin, the intensity of the diffraction pattern will have an oscillatory component. That means two atoms of the same speciens might appear in opposit contrast.
                
                %

             \item Annular Dark-Field Imaging
             
             The detector is ring shaped. It integrates the scattered electrons, excluding the primary beam. 
             
            
             An important example of ADF is the high angle annular dark field HAADF mode, which is
             suitable for imaging materials which have high atomic number. The resulting image has high contrast, with the scattering cross section being a function of $Z^2$. The resolution is dependent on signal to noise ratio and the beam diameter.\citep{stemb}
             
            
            
            
        \end{itemize}
    
    
    STEM has other operating modes such as:
    
    \begin{enumerate}
        
        
        
        
        
        
        
        \item Annular Bright Field ABF
        
        In this mode, the image is formed by integrating the intensity of the scattered electrons near the primary beam using a ring-shaped detector. 
        
        This mode enhances the contrast for light atoms in STEM. 
        It is capable of characterizing materials whose crystals contain light and heavy elements.
        
        \item Differential Phase Contrast DPC
        
        The gradient of the object potential along the direction of the scan determines the image contrast. This mode is suitable for studying e.g. magnetic domain structures and electronic fields.
    \end{enumerate}

        
        
        
        
        

        \begin{figure}[H]
        \centering
        \includegraphics[scale=0.29]{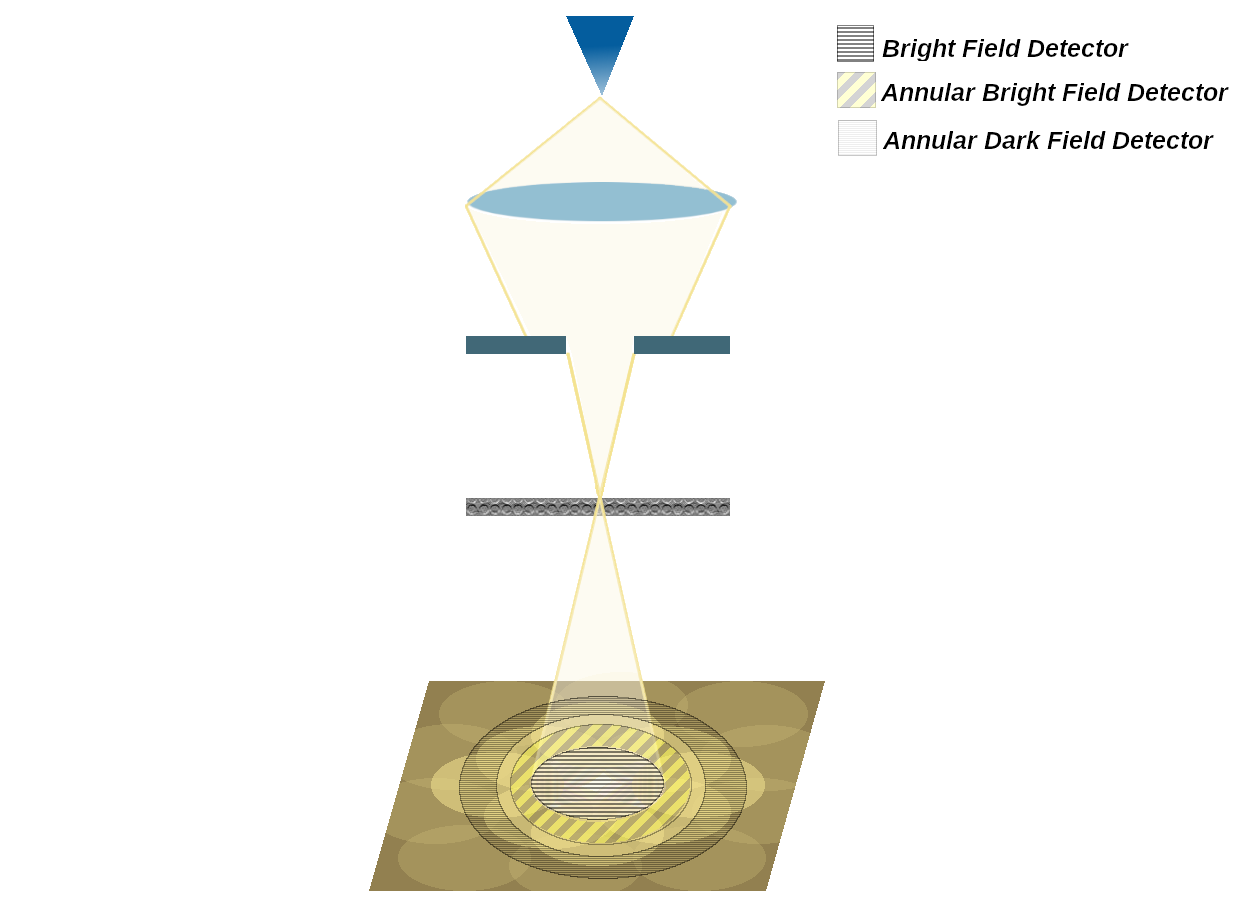}
        \caption{In the conventional imaging modes of STEM the intensity of the diffraction pattern is integrated over the region of the specific detector. The result is two dimensional image in real space. The diagram shows the regions covered the bright field (1), annular bright field (2) and annular dark field (3).}
        \end{figure}
   
        In the following, the modes mentioned above will be refered to as the conventional STEM modes. For prychographic recostruction of the STEM image, a special detector is used. It collects all the information available in the field of view. The mathematical method of ptychography is then used computationally to reproduce the phase and amplitude of the spacimen , which is known as the object transfere function (OTF). Several methods of ptycography allow the retrieval of the the probe function as well. This mode is known as the momentum resolved STEM or 4D STEM.


   \section{Retrieving the phase and amplitude of the Object Transfer Function (OTF) using ptychographic and non-ptychographic solutions}

\begin{figure}[H]
\centering
\includegraphics[scale=.4]{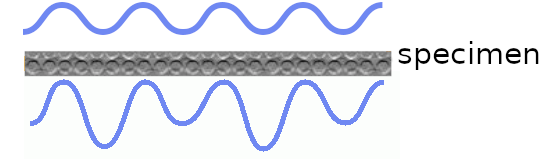}
\caption{Electron waves passing through a specimen will suffer modification in both phase and amplitude}
\end{figure}

In STEM the Fraunhofer far field diffraction pattern of the object is recorded in the back focal plane of a post-specimen lens. Roughly speaking, the transmittance of the object is given by the modulus of its exit wave. While its phase is the accumulated phase difference, relative to free space. The phase difference is caused by the thickness of the object, in particular the real part of its refractive index.

Jumping between the object function and the diffraction pattern is possible using Fourier Transform. While it is straightforward to find the diffraction pattern from the object function, the inverse process is not unique. In particular the phase of the  diffraction pattern is 
unknown because what is recorded is the intensity which is the square of the amplitude of the electron wave function at the diffraction position. 
Therefore, the OTF phase is intractable. 

Recall that what is recorded is the intensity, and while the wave amplitudes add linearly, their intensities don't. Therefore, the resulting solution space is highly non-linear.

For transparent objects, like biological cells for example, the modulus of their exit wave is approximately 1. They majorly introduce  phase modification to the waves passing through them. From this we understand the
pressing need for phase retrieval over all wavelengths. 

While this is an important aspect of the success of high resolution x-ray ptychography, it is the main motive for both visible light and electron ptychography. 
It must be mentioned that 
 the following sections 
mainly follow \cite{pty2019}.

\subsection{Non-ptychographycial Iterative Solutions Method for the OTF problem}

    The support of the object is the finite 2 dimensional area that delineate the object. Each Fourier component of the object exit wave corresponds to a single  pixel in the diffraction pattern. Changing the phase in a 
    scan position 
    is equivalent to a lateral shift in the Fourier component of  the object function with the corresponding frequency (one to one map).
    
    To find a unique solution, an iterative computational loop is set up such that part of an estimated object function is selected by the well known aperture. The two components are compared to the measured data and its estimated modulus and phase using Fourier/Fresnel propagation and back propagation transforms.
    
    \textbf{\large{Constrains}}
    
    \begin{itemize} 
        \item The modulus of the wave function is the square root of the intensity of the data.
        
        \item No amplitude exists outside the extent of the aperture.
        
        \item The loop has an arbitrary starting point unless extra a-priory knowledge is given. For example, knowing the object is transparent, it is convenient to start the loop from an image field in the diffraction space with total transparency in every image pixel.
    
    \item The loop might take up to 10000 iterations to converge.
    
    \item This approach has dozens of variations
    
    \end{itemize}

    \subsection{Ptychography: multiple diffraction pattern}
    
    The detector is of a fixed size. A much bigger matrix is computationally reserved for the object, each pixel (element) of which matches the detector size. The illumination is  raster-scanneded across the specimen (or vise versa), recording a diffraction pattern at each position. An iteration loop is applied each time.
    
    \begin{figure}[H]
    \centering
    \label{diffpatt}
    \includegraphics[scale=.23]{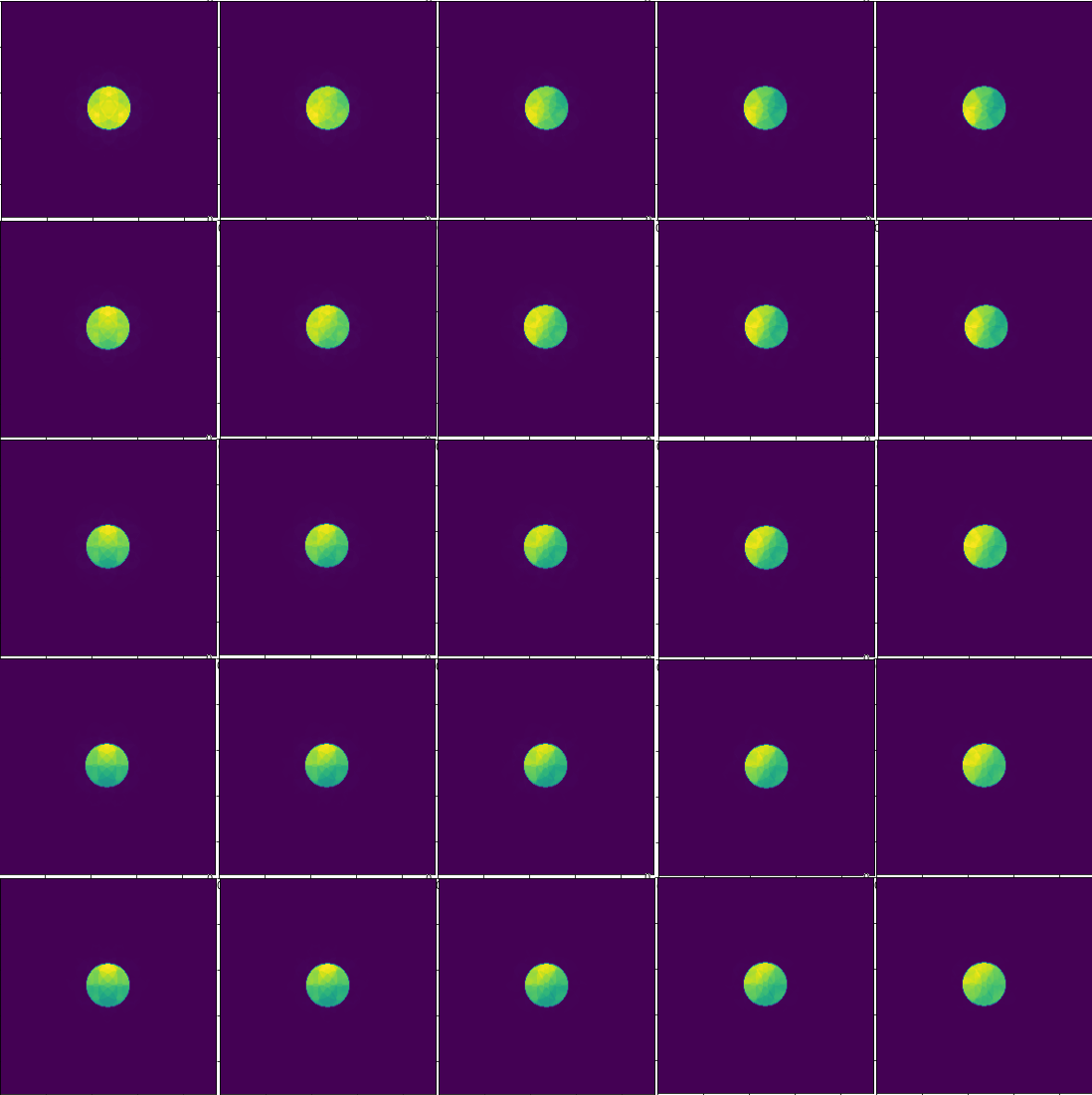}
    \caption{A subset of the 4D dataset. At each position in real space a diffraction pattern is recorded. The transmitted beam has much higher intensity than the scattered electron beams which cover the whole detector area.}
    \end{figure}
    
    Knowledge from the experiment constrain the problem. The 2D finite delineated area of the specimen is called the support area of the specimen.  Each pixel in the diffraction pattern is a single component of the Fourier transform of the object exit wave. Laterally shifting the fourier wave component corresponding to a certain frequency is equivalent to changing the phase of the equivalent pixel. 
      Knowledge from the experiment reduces the set of possible phases. Unavoidable ambiguities include the shift of the whole object function or that the diffraction pattern is the same for both the object and its Hermitian conjugate.

    
    \textbf{\large{Advantages}}
    
    Ptychography is robust to noise and imperfections of the optical setup. The resolution of the resulting image exceeds the information limit. 
    The only limitation to ptychography resolution is the electron wavelength.  Here is a list of advatages of Ptychography over other methods;
    
    \begin{itemize}
    
        \item It provides prior knowledge of the illumination (relative) positions.
        
        \item Measurements provided exceed the unknowns. Some of the unknowns are expressed in more than one diffraction pattern.
        
        \item The subset of the object function where the object wave must be identical in both diffraction patterns, under the given conditions, is reduced drastically.
    \end{itemize}

    \subsection{Requirements for Ptychography}
    
    In order to retrieve the object function using ptychography, certain conditions must be satisfied:
    
    \begin{enumerate}
        
        \item Overlapping of the illuminated area
        
        If the area covered by the aperture are mutually exclusive, some of the ambiguities of the phase problem will remain \cite{fr}. The processing time when using overlapping areas is much faster. Ambiguities, especially centrosymmetric ones, are then avoided.
        
        \item The computationally reserved matrix for the object function must be (infinitely) bigger than the detector size. Each pixel in this matrix has the same size as the detector.
        
        \item Multiple diffraction pixels.
        
        A single diffraction pixel represents the lowest sampling limit in diffraction space and very dense sampling in real space. However, it does not solve the phase problem.
        
        The minimum requirement for ptychography is doubling the information in x and y dimensions by using 4 pixels detector (or 4 quadrants in a circle).

    \end{enumerate}

\subsection{Examples of Ptychographical 
methods}

Ptychography is applied to to STEM data as well as TEM , x-rays and visible light. 

\begin{enumerate}
    \item Fourier Ptychography
    
        The resulting image is equivalent to conventional STEM ptychography. Here a TEM set up is used with an aperture placed in the back focal plane. The lateral displacement in illumination position is replaced by the tilt of the incidence angle  of the electron plane waves. Then a single pixel in the image plane is equivalent to a diffraction pattern when mapped against all tilts. 
      
    \item Selected Area Ptychography
    
    A method used in TEM for characterizing small areas of the specimen for example small area grain or small isolated object. This technique can image a very large field of view and can map the electric or magnetic fields. Nevertheless, it is most extensively used with visible light in characterizing biological cell life cycle
    (\cite{biopty1}, \cite{biopty2}). 
    
    A TEM set up is used with an aperture placed on the image plane. The specimen is displaced laterally with respect to the aperture. The resulting diffraction pattern at a further plane down the optical axis is recorded. Here the OTF is the CTEM image and the aperture is the illumination. 
    
    
\end{enumerate}

\section{Ptychographis methods used in STEM}

Focused-probe methods of ptychography are used  for OTF reconstruction in STEMs. They are  divided into iterative and non-iterative methods.  

For iterative methods, the probe function is updated using forward and backward propagation of the localized object and the measured diffraction pattern.

Mathematically, ptychography tries to find the intersection between the set of the OTF solutions, which satisfies the recorded diffraction pattern and the set which satisfies the real space priors i.e. the aperture.

Given the space of all possible solutions of the OTF problem, the unique solution, if exists, is the intersection point between the set of the diffraction pattern intensity and the set of the real space priors. 
This solution maybe found using 
a projection algorithm which alternatively projects an estimate solution point in one set to the nearest point in the other. If the two sets are convex, the loop will definitely converge to the correct solution. However, the phase problem is not convex, the loop can possibly get stuck at a different point.

A better method would employ projection and reflection. Examples of this method are the difference map (DM) and the relaxed averaged alternating reflections (RAAR). Such methods speed up the convergence, widen the accessible search space and avoid local minima.

For the non-iterative, direct ptychographical methods for real space sampling,
    %
     very densely sampled data are used in  linear 
     inversion methods to solve the ptychographic problem. Such methods are equivalent to the modern iterative ones regarding the 
     aspect that it solves for the illumination and removes the potential coherence effects.
     %
        
            
            
            
      %
    while having the advantage of being computationally very fast. These methods include
    \begin{enumerate}

        \item Single Side-band    

        \item Wigner Distribution Deconvolution.
    \end{enumerate}

    In the following sections a thorough description is given to some of the commonly used iterative and non-iterative ptychographic algorithms.


\subsection{Dense Sampling}

Before giving a thorough example of non-iterative ptychography, it is important to understand the concept of dense sampling.

The interfering diffracted orders make fringes perpendicular to the scattering vector of the diffracted reflections. The periodicity of these fringes is identical to  the featured cast in the shadow image of the object function. 

\begin{figure}[H]
\centering
\includegraphics[scale=.7]{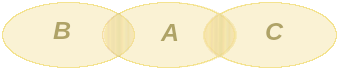}
\caption{The interference diffracted orders' fringes has the same periodicity of the featured cast in the shadow image of the object function.}
\end{figure}

For coherent imaging; the isolated features of the specimen are not easily visible. 
%
%
The periodicity of the specimen determines the position and features of the diffracted beams. As the probe moves across the specimen, the features within the overlapping areas change with the same periodicity. 
If the probe is focused, no features appear on the overlapping disks.  
The solution of the phase problem, comes from the relative positions of these fringes. 
%
%
The phase  difference between each pair of diffracted discs is given by the phase of the  features of the Fourier transform w.r.t the probe position of the intensity function. Influences of aberrated 
 defoucused probe might still, however, need to be corrected / deconvolved.





\textbf{Sampling limit}

Having a maximum sampling in real space implies tolerance of minimum sampling in reciprocal space if the object is weakly scattering. However, minor aberrations will then not be removable and the resulting resolution can not be better than the one defined by the lens aperture.
The maximum sampling in the reciprocal space is given by the size of the detector pixel being less than the reciprocal of the whole field of view  of the reconstructed image, rather than the inverse of the size of the probe.
For weak phase objects only certain points of high interest are important. 
In the unrealistic case of perfect crystal with only one overlap is present at any point in the diffraction pattern, the phase problem can be solved with only two positions.

The very densely sampled data are used in a linear non-iterative inversion method, to solve the ptychographic problem. Such a method is equivalent to the modern iterative ones regarding the following aspects
    \begin{itemize}
    
        \item Solves for the illumination,
        
        \item Remove partial coherence effects,
        
        \item Suppress 3-D scattering effect,
        
    \end{itemize}
    
while having the advantage of being computationally very fast.


    \section{The need to  go beyond conventional STEM}

    Using ptychography allows phase recovery and super-resolution by using all the available information in the field of view.
    The principle of STEM as ptychograph was introduced by \citep{hoppe1}\citep{hoppe82}. \citep{tsr} have mathematically demonstrated the Wigner Distribution Deconvolution (WDD) phase retrieval method, already pointing out its high potential in producing high-resolution, robust and noise-free images.
    Even at early implementation stages of the Wigner distribution deconvolution reconstruction method on a weak phase object, which is known as the single side-band ptychography (SSB) by \citep{exp93}, a resolution twice higher than the one permitted by the aperture has been achieved. While the conventional resolution permitted by the setup was 0.93 nm the resolution achieved was 0.46 nm. 
        
        \begin{figure}[H]
        \centering
        \includegraphics[scale=0.35]{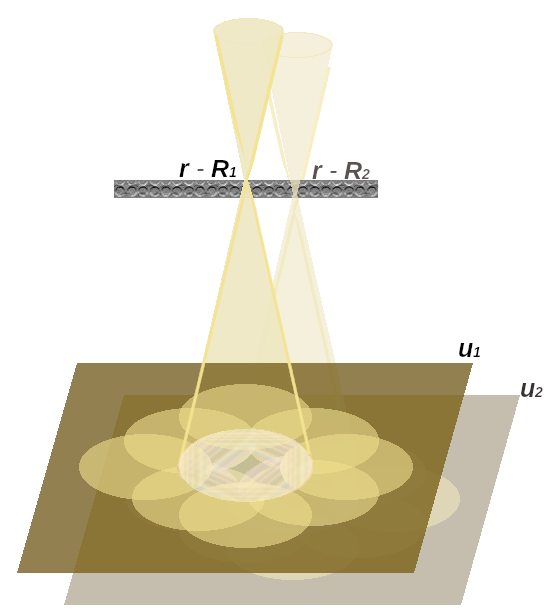}
        \caption{The momentum resolved data recorded for the ptychographic reconstruction is shown. At each scan position of the illumination ($\vec{r} - \vec{R}$), the two dimensional  diffraction pattern  ($\vec{u}$) is recorded. The resulting data is 4 dimensional; two dimensions in momentum space (the diffraction pattern) and two dimensions in real space (the scan position). This method is also known as 4D STEM}
        \end{figure}

   \section{Implementations of 
   Ptychography }
   
       
       

       
       
       Ptychography has shown high resolution imaging that is robust to noise. In the following,  examples of STEM ptychography implementations are explored with emphasise  on the strength points they demonstrated.
       
       \begin{itemize}
       \item Double resolution:
       
       At early implementation of the WDD ptychographic reconstruction, double resolution to what is typically achievable by the optical setup has been reached \citep{exp93}. The self consistancy of ptychography has also been proven against various resolution degrading effects.

       \begin{figure}[ht]
       
       \vspace*{1cm}
       \centering
       \includegraphics[scale=0.28]{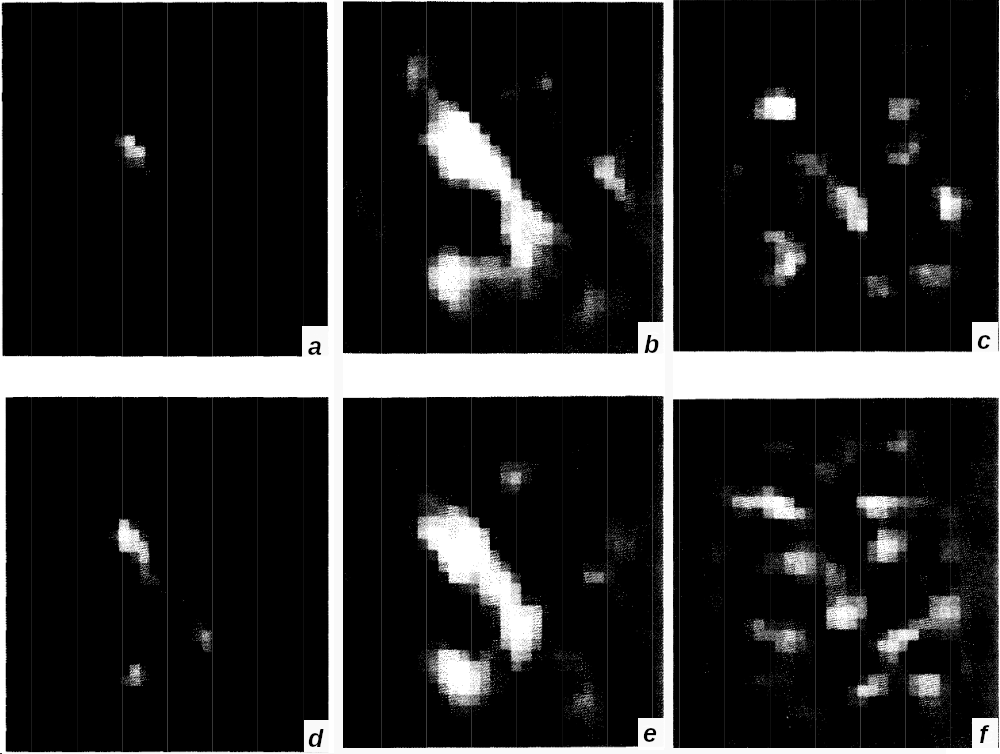}
       \caption{The ptycographic reconstruction of the amplitudes (b) and (e) and the phases (c) and (f) is found using the simplified method of the Wigner distribution deconvolution, similar to what is now known as the single side-band. The corresponding conventional STEM bright field images are shown in (a) and (d). The nominal resolution given by the setup is 0.93 nm while the resolution of the prychographic reconstruction is 0.47 nm. The two data sets are of the same amorphous specimen, recorded consecutively \citep{exp93}.}
       \end{figure}

       \item Resilience to corrupt data:

            \begin{figure}[ht]
                \centering
                \includegraphics[scale = 0.37]{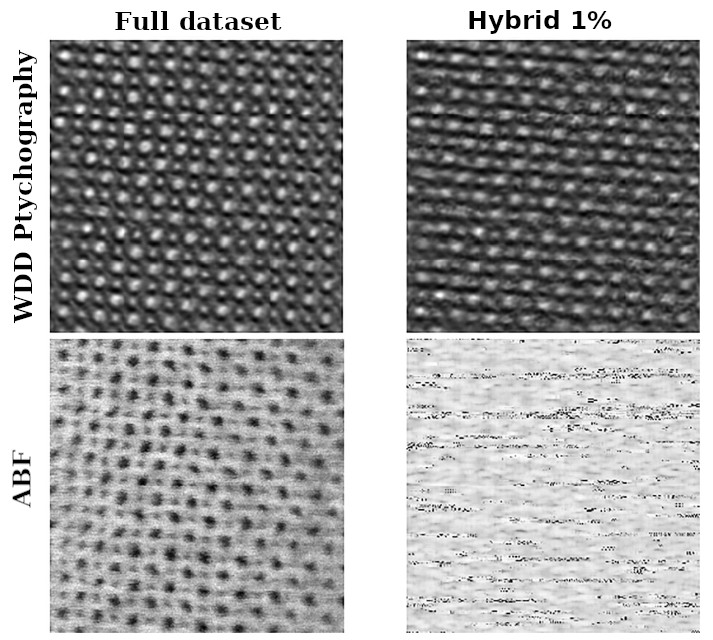}
                \caption{The hybrid 1\% image correspond to the case in which there are missing parts of the 4D dataset that are poorly recovered.
                While the WDD ptychographic method succeeds in retrieving a usable phase-image in the case of corrupt data (up-right), the ABF image of the corrupted data (down-right) does not present any information about the specimen and incomparable to the original image \citep{ssstempty}.}
                \label{wddcrptn}
                \end{figure}

           \citep{ssstempty} have shown that the WDD reconstruction is empirically resilient to corrupt data. A comparison of a very poor 
           data set which has several artifacts including 
           missing data, has been made between WDD ptychographic phase image and ABF image. The results are shown in figure \ref{wddcrptn} with comparison to the original images. 

       \item High dose efficiency:
       
           Dose efficiency is an important aspect of imaging applications with electrons, x-rays and neutrons. For limited dose imaging, the goal is to make the most of the signal available before the specimen is damaged or destroyed. For atomic resolution imaging, the suitable methods that achieve such high conditions are Conventional TEM (CTEM) and STEM. While STEM is popular for material characterizations due to its analytical properties, high resolution CTEM (HRTEM)
           is currently favoured for biological specimens due to its phase contrast imaging. 
           \citep{hde19} have shown that using ptychographic reconstruction algorithms produces readily interpretable higher resolution images. SSB ptychography has been used on simulated crystalline as well as biological structures at atomic resolution. The ptychographic STEM phase images have shown 
           superior dose-efficiency to conventional HRTEM for light materials. 
           under low dose conditions.
           
           \begin{figure}[H]
           \centering
           \includegraphics[scale=0.5]{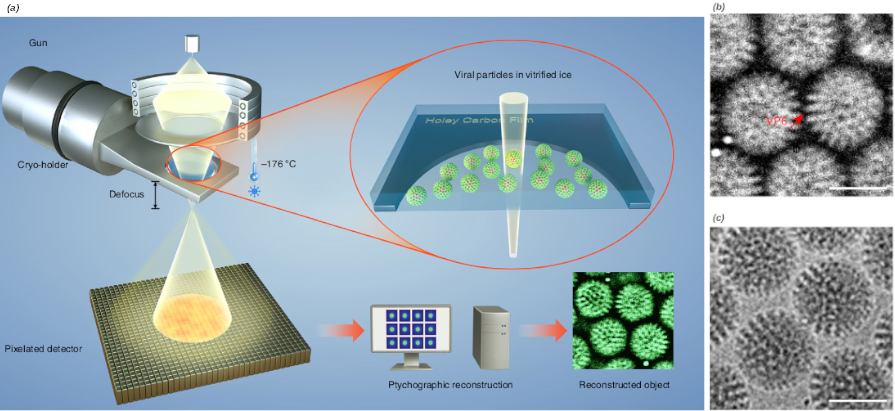}
           \caption{(a) shows a schematic of the setup and the reconstruction method. To reduce the dose on the biological specimen, high defocus was used as shown on the amplified section. Ptychographic reconstruction using ePIE was used to reconstruct the image of the specimen shown on the lower corner. This experiment demonstrates the high potential STEM ptychography has in biological imaging under high defocus and low dose conditions. The experiment was conducted in a cryogenic environment at T= 176 C.  Phase images are shown in (b) and (c). The dose in (b) is 22.8    $e^-$ / \AA$^{2}$ The dose in (b) is 22.8    $e^-$ / \AA$^{2}$ and in (c) is 35 $e^-$ / \AA$^{2}$. The scale bar is 50 nm.\citep{dose}}
           \label{dosefig}
           \end{figure}

           Recent studies \citep{prufbsd20} and \citep{dose} have demonstrated the potential of ptychography  as a low-dose 4D STEM technique. The high defocus technique and its corresponding ptychographic results are shown in figure \ref{dosefig}. 

       \item Aperiodic specimens
        
            SSB ptychography has been applied to simulated data of  non periodic specimen under partial coherence conditions. High resolution  phase images have been obtained by \citep{hde19}.
            
            \begin{figure}[H]
            \centering
            \includegraphics[scale = 0.5,trim = 9cm 11.5cm 16cm 8cm,clip]{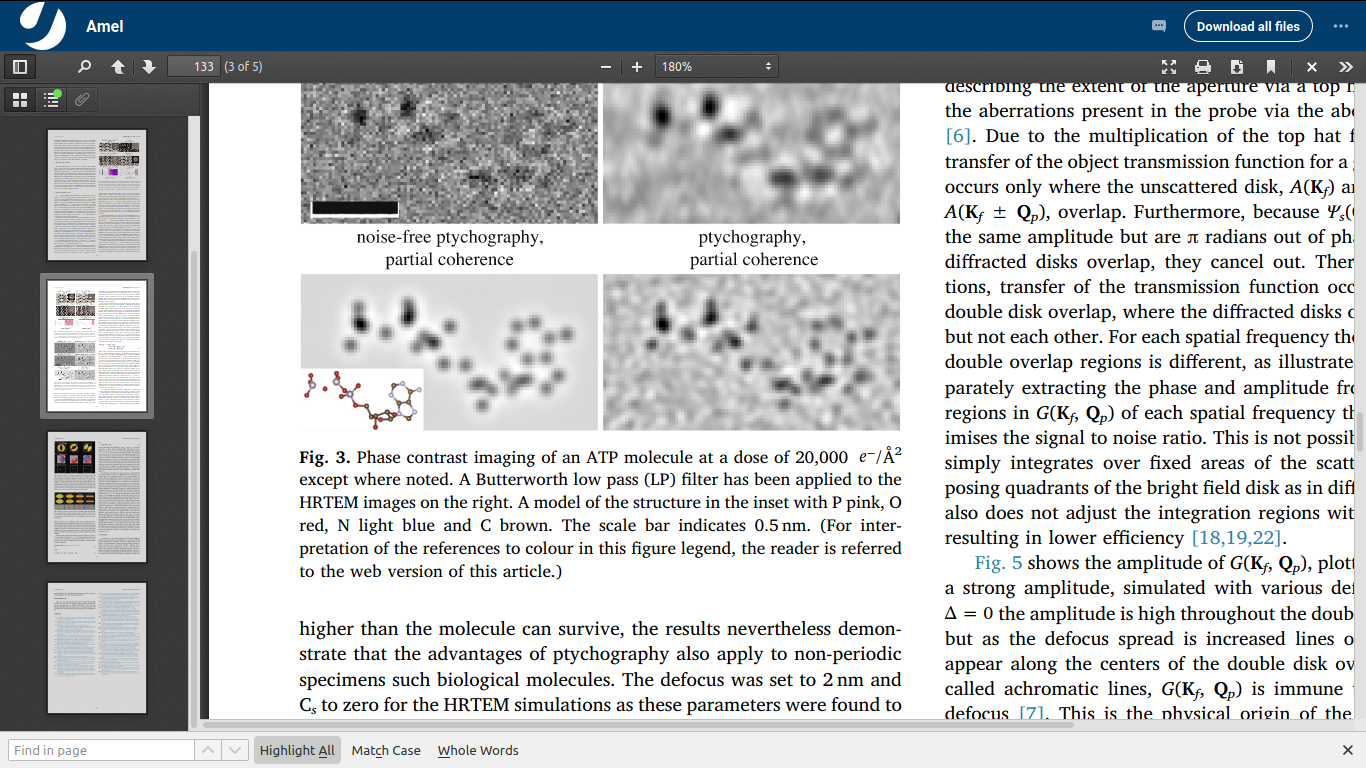}
            \caption{Phase contrast imaging of simulated data of an adenosine triphosphate (ATP) molecule at a dose of 20,000 $e^-$ / \AA$^{2}$. The model structure at the corner shows P in pink, O in red, N in light blue and C in brown. The reconstruction is made using SSB method. \citep{hde19}. }
            \label{nonperiodic}
            \end{figure}

        \item Reducing the acquisition time:
       
           High resolution STEM imaging with pixelated detectors is much slower than conventional STEM imaging methods.
           In early implementations of STEM ptychography in the 90s, the data of 64 x 64 detector pixels x 32 x 32 scan positions took 80s \citep{exp93}.
           Nowadays the 4D data sets are typically of the size (256)$^4$. 
            This problem was typically treated by using highly defocused probe to reduce the number of diffraction patterns collected, meanwhile sufficient overlapping of the illumination on the specimen is ensured \citep{aipp09} \citep{pamhadf12} \citep{dep14}. 
            The adaptation of fast cameras 
           allowed the use of sub-{\AA}ngstr\"{o}m electron probe. As a result, acquisition of incoherent Z-contrast imaging and spectroscopic signals can be made simultaneously with the  ptychographic diffraction patterns. 
            %
            This combination allows 
            the strong compositionsal sensitivity to element type of the z-contrast imaging and the efficient quantitative phase imaging of ptychography. For example a full determination of a previously unknown carbon nanostructure consisting of both heavy and light elements has been made under this combination as shown in figure \ref{zpty} \citep{Yang2016}.
            
            \begin{figure}[H]
                \centering
                \includegraphics[scale= 0.37, trim = 3cm 3.5cm 4cm 7cm, clip]{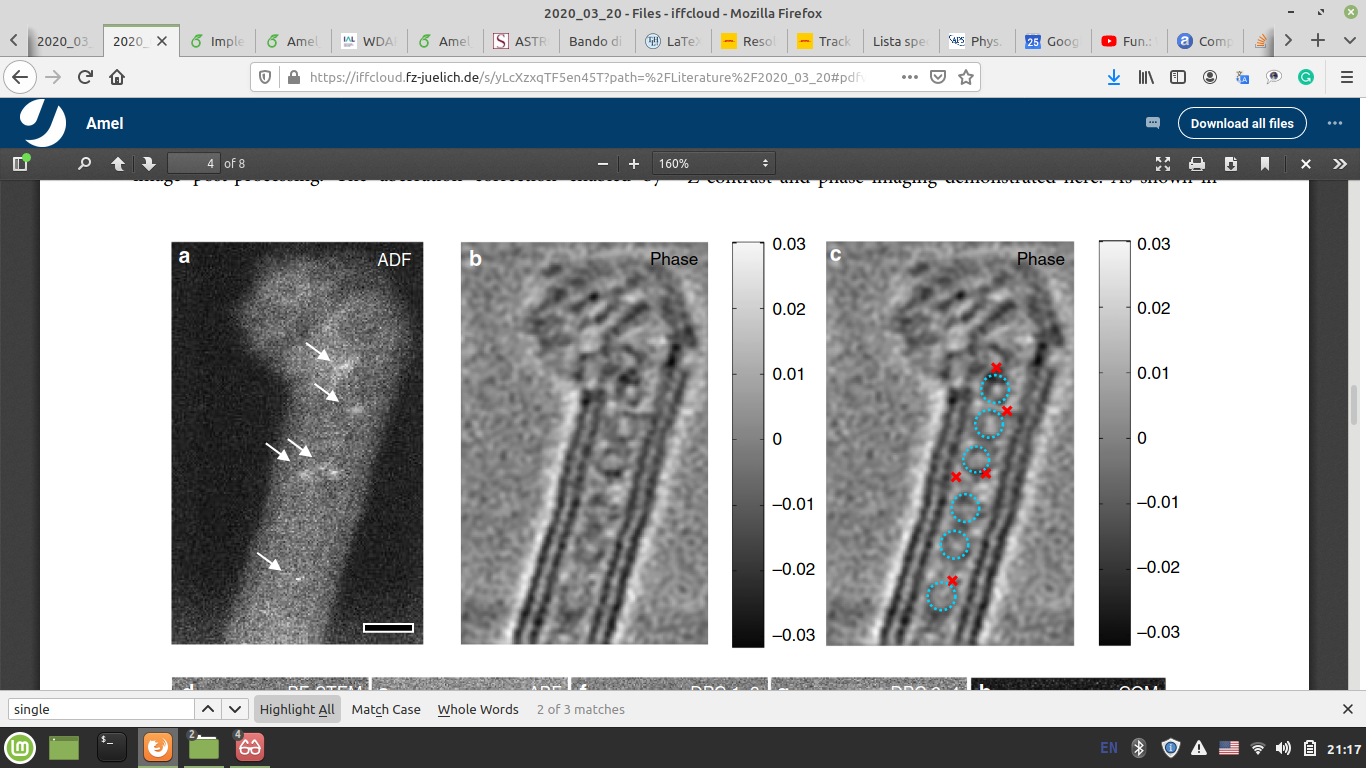}
                \caption{Characterization of Double-wall complex carbon nanotube (CNT) peapod. (a) Z-contrast ADF image. The arrows indicate the location of the single iodine atoms. (b) shows the phase image reconstructed using WDD method. In (c) the positions of the fullerenes is annotated on the phase image using the dotted cycles and of the iodine with the x marks. The positions are determined based on the corresponding location on the ADF image \citep{Yang2016}}
                \label{zpty}
            \end{figure}

        Nevertheless, the speed of those cameras, at the speed of $10^3$ frame per second (fps), is still 3 orders of magnitude lower than the scanning speed of a conventional STEM. 
            %
            %
            On the other hand, sub-sampled imaging have been proven to reduce both dose and acquisition time. Recently \citep{ssstempty} have shown that by randomly removing both detector pixels and image position, faster data collection is achieved. They used WDD algorithm to obtain the phase image. This method reduced the acquisition time by a factor of 100, making STEM-ptychography competitive with conventional STEM. It has been shown quantitatively that the resulting phase-images under low levels of sub-sampling (25\%) were almost identical to full sampling ones. Meanwhile, under very low sub-sampling (1\%) useful phase-images have still been produced.
            
            Previously, machine learning methods for sub-sampling which employ compressive sensing algorithms had been proven to be efficient for high-resolution STEM imaging under extremely low dose conditions ($\leq 1 e^-$ / {\AA}$^{2}$) \citep{mlpaperdose} \citep{phdml18}.  The deliberate sub-sampling was at least an order of magnitude faster than the conventional low-dose method. Recovery algorithms have then been used to decompress the compressed acquisitions. The ptychographic recovery algorithms have been shown to be robust to either scan position or detector's low-resolution \citep{daSilva}. 
           
           A combination of a Fourier filtering technique \citep{daSilva} and WDD ptychography has been implemented by \citep{ssstempty}. The Fourier filtering technique can reduce the reconstruction time to seconds. It already integrate some of the WDD steps required. This method can potentially be used for real-time phase retrieval.
           
           Recent development in STEM detection parameters has been able to achieve 7500 frames per second, allowing less than 40 s total scan times for a 512 x 512 STEM image \citep{camnellist19}. 
           Phase reconstruction of an aluminosilicatezeolite (ZSM-5)  has been made using SSB ptychography.  
           More recently \citep{prufbsd20} recorded only 0 or 1 at each detector pixel. This method has allowed to decrease readout time of the on-chip counters (the dwell time) and increase the frame rate to 12.5 kHz.
           
           A good matching between the camera and scanning speeds will allow new applications in both materials science and biology. For example, the possibility of in-situ dynamical STEM ptychography.
           
           \begin{figure}[H]
            \centering
            \includegraphics[scale = 0.37,trim = 0cm 4cm 1cm 13cm,clip]{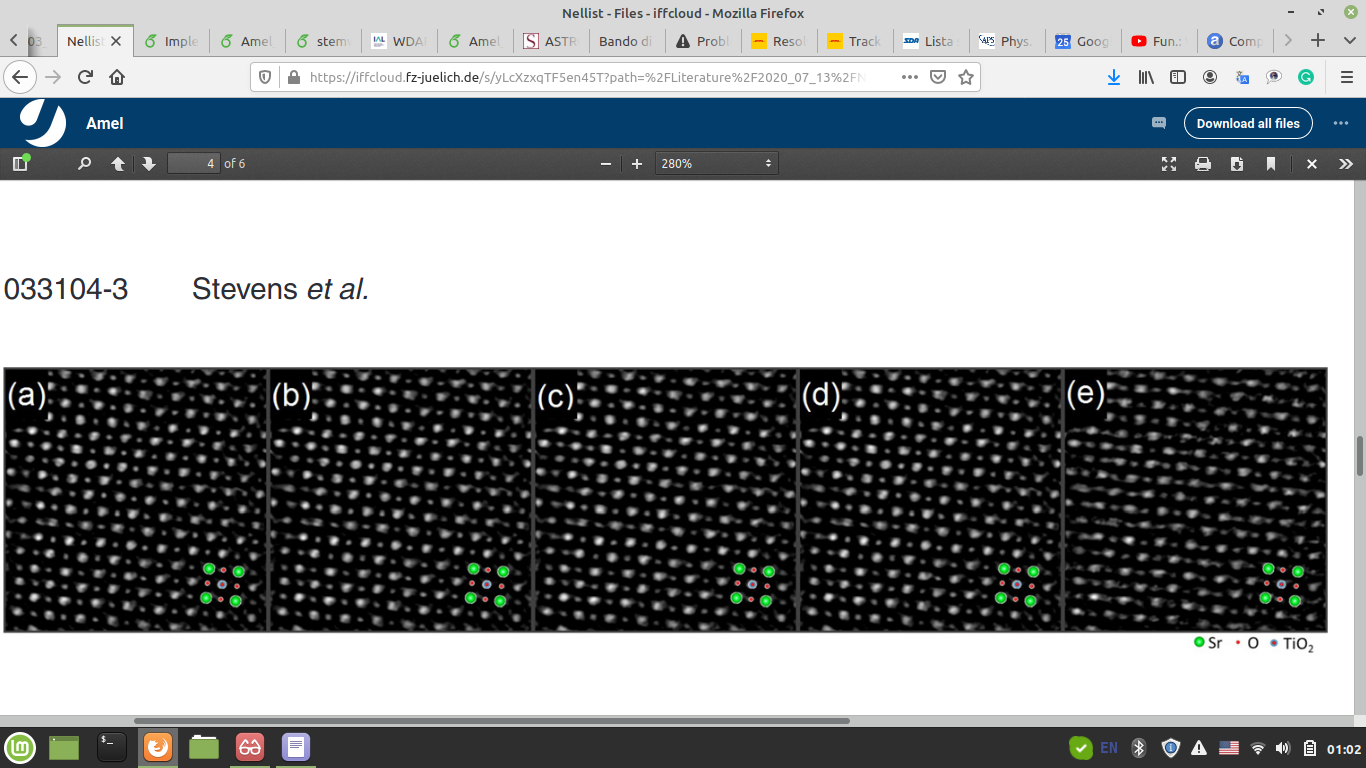}
            \caption{Ptychographic phase-images of a SrTiO3 wedge: (a) original, (b) 25\% detector sub-sampling, (c) 25\% probe sub-sampling, (d) 25\% hybrid sub-sampling of 50\% detector and 50\% probe, and (e) hybrid sub-sampling of 10\% detector and 10\% probe. For better visualization, the image intensities are scaled by their maximum intensity and all negative values are set to 0 (i.e., black). The WDD method was used for the recovery. \citep{ssstempty}.}
            \label{nellist subsamble}
            \end{figure}

       \end{itemize}

       
    \section{Outline of the thesis}
    
    In the second shapter the mathematical derivation of certain STEM ptychographic algorithms is given starting from the discribtion of the electrom wave interactions through the STEM column. 
    
    The final chapter presents the results I obtained during the thesis.
    
%
%

%% file: chapter1new.tex

\chapter{Ptychographic reconstruction of STEM images}

This chapter concerns the STEM image reconstruction using ptychography. It consists of two main parts. First, the formation of the diffraction pattern at the detector position of the STEM, which is disscussed in the sections 1 to 5. The ptychographic reconstruction of the image is thereafter discussed in the rest of the chapter.

The general conceptual background about the microscope resolution and electron wave is given in section 1. Then interaction between the electrons and the specimen and the probe formation are disscussed seperatly in sections 2 and 3. As Fourier transform and convolution are the main pillars of the mathematical description given in 2.3 and 3.3, the concept of convolution is thereafter disscussed mathimatically. Leading finally to the description of the differaction pattern at the detector.This part Generally follows \citep{ictem}.

The princible of ptychography is then discussed in sections 5 and 6.   Examples of iterative and direct ptychographic methods are given. 
This part generally  follows \cite{pty2019}.


In the next section,  the electron wave interactions along the microscope column will be discussed.   
 The two most important processes in transmission electron microscopy are the scattering of the probe wave from the object and the subsequent interference between the wavefronts at the detector position. 
 This is what is about to be discussed.


\section{Funamental princibles}

The 
 resolution
 $d$ of an image formed using a microscope is limited by the wavelength of the illumination. This is given by the Abbe resolution limit \citep{abbe}
 
  \[d > \lambda,\] 
  
  Thus according to this diffraction limit, in order to obtain atomic resolution ($d \sim \AA$) one would need wavelengths comparable to the atomic distances, which are considerably smaller than the wavelenvgth of visible light ($\lambda \sim 10^2 nm$). 
%
 Accelerated electrons then present a suitable media for such fine imaging. At acceleration $E=300$ kV, the electrons have wavelength $\lambda = 1.969$ pm and they travel  at about 0.78 times the speed of light. It is therefore important to take relativistic considerations into account.
       The physics of the system is therefore governed by Helmholtz equation 
        
        \begin{equation}
        \label{rele}
        \Delta \psi + 4 \pi^2 u_0^2 \psi = 0.
        \end{equation}

         Where $\Delta$ is the Laplacian operator and $u_0$ is the wave vector of the electron waves. The electron wavelength is therefore given by; 
        
         \[
          \lambda = \frac{h}{\sqrt{2meE+\frac{e^2 E^2}{c^2}}}
          %
         \]
        
        
        The condition for the electrons to form  constructive interference after transmitting throught the specimen is that the path difference between the unscattered and the Bragg-scattered electrons must be an integer multiple of the wave length 
        
        \[
        2 \frac{d_{hkl}}{n} sin\theta = \lambda
        \]
        
        Where $hkl$ are the Miller indices specifying the plane at which the electron is scattered.

    \section{The interaction between the specimen and the illumination}
    
        \subsection{Elastic scattering of electrons by an infinite crystal
        }
    
            
            An infinite lattice can be represented by a set of delta functions located at periodic positions corresponding to the  unit cells.
            
            \begin{equation}
            \mathcal{T}(\vec{r})
            = 
            \sum_{a,b,c=-\infty}^{\infty}
            \delta
            (\vec{r} - \vec{t}_{abc})
            .
            \end{equation}
            
            where $\vec{t}_{abc}$ is a lattice vector with coefficients $a,b,c$. Each unit cell consists of $N$ atoms. Therefore, the potential of the cell is given by;
            
            \begin{equation}
            V_{cell}
            (\vec{r})
            = 
            \sum_{j=1}^{N}
            V^a_j
            (\vec{r} - \vec{r}_j)
            .
            \end{equation}
            
            The complete potential can then be expressed as a convolution between the set of the delta functions which represents the lattice and the unit cell potential.
            
            \begin{equation}
            V(\vec{r})
            = 
            V_{cell}
            \otimes
            \mathcal{T}(\vec{r})
            .
            \end{equation}
            
            Using the periodicity of Bravais' lattice, the plane wave expansion can be used to express the crystal potential.
            
            \[V(\vec{r})
            = 
            \sum_{\vec{g}}
            V_{\vec{g}}
            e^{2 \pi i  \vec{g}  \vec{r}}
            .
            \]
            
            Where $V_{\vec{g}}$ are the Fourier coefficients corresponding to each lattice plane in the real space.
            
            Given that $V (\vec{r})$ is real, and if the reference frame is located at an inversion center, then $V_{-\vec{g}} = V_{\vec{g}}$. Therefore the Fourier Transform of the potential is 
            
            \[V (\vec{u})
            =
            \frac{1}{\Omega}
            \int \int \int
            V (\vec{r})
            e^{-2 \pi i  \vec{g}  \vec{r}}
            d\vec{r}
            =
            \frac{1}{\Omega}
            \mathcal{F}[
            V (\vec{r})]
            =
            \frac{1}{\Omega}
            \mathcal{F}[
            V_{cell} (\vec{r})]
            \mathcal{F}[
            \mathcal{T}
             (\vec{r})]
            .
            \]
            
            Where $\Omega$ is the volume of the unit cell.
            
            However,
            
            \begin{align}
            &\frac{1}{\Omega}
            \mathcal{F}[
            V_{cell} (\vec{r})]
            = 
            \frac{1}{\Omega}
            \sum_{j=1}^{N}
            f_{j}^{e}
            e^{-2 \pi i  \vec{g}  \vec{r}}
            \equiv
            V_{\vec{g}}
            ,
            \label{vg}
            \\
            &\mathcal{F}[
            \mathcal{T}
             (\vec{r})]
            = 
            \sum_{\vec{g}}
            \delta
            (\vec{u} - \vec{g})
            .
            \end{align}
             
             Implying that
            
            \begin{equation}
            V(\vec{u})
            = 
            \sum_{\vec{g}}
            V_{\vec{g}}
            \delta
            (\vec{u} - \vec{g})
            .
            \end{equation}
            
            This expression shows that the Fourier Transform of infinite crystal is a discrete function with non-zero values only at the reciprocal lattice points $\vec{g}$. 

        \subsection{The case of a finite crystal}
        
            To account for the finite size of the crystal,  the infinite crystal potential is multiplied by the shape function $D(\vec{r})$, to yield the finite crystal potential
            
            \begin{equation}
            V_f(\vec{r})
            =
            V(\vec{r})
            D(\vec{r})
            .
            \end{equation}
            
            Where,
            
            \[
            D(\vec{r})
            =
            \left\{
            \begin{array}{lr}
            1 &  \text{inside the crystal}\\
            0 &  \text{outside}.
          \end{array}
            \right.
            \]
            
            The Fourier transform of the potential is then found using the multiplication theorem;
            
            \[
            V_f(\vec{u})
            = 
            \mathcal{F}[
            V(\vec{r})]
            \otimes
            \mathcal{F}[
            D(\vec{r})]
            .
            \]
            Yielding;
        
            \begin{equation}
                V_f(\vec{u})
                = 
                \sum_{\vec{g}}
                V_{\vec{g}}
                D(\vec{u} - \vec{g})
                .
            \end{equation}
            
            This expression means that the lattice points in the reciprocal space have a certain shape determined by the shape of the crystal in the real space. 
            Notice that unlike the case for the infinite crystal, the Fourier coefficients are continuous functions of the reciprocal space vector $\vec{u}$ . 

            
            
            
            
            $V_g$  defined in \eqref{vg} is known as the structure Factor. 
            If the specimen is crystalline; there will be only few values of $\vec{u}$ at which the value of $f_j^e$ is significantly larger than its neighboring values. 
            These maximum values correspond to the Bragg reflections for this particular projection of the crystal. In between the Bragg reflections, the value of these Fourier coefficients is zero, in the absence of thermal diffuse scattering \citep{acem}.
            
            
            
            A correct theory for thicker specimen and/or orientations close to Bragg's must allow multiple scattering for each electron. Therefore, the kinematical approximation, where only first bragg scattering is allowed, is not applicable in the case of electrons.
            

        \subsection{The dynamical scattering of the electrons by the specimen}
        
        As electrons travel through the crystal they first suffer refraction. That is the acceleration associated with  the zeroth component $V_0$ of the Fourier transform of the complex potential $V_c(\vec{r})$ function of the lattice. Here the potential is presented to be complex because the inelastic scattering is taken into account. The inelastic scattering potential $W(\vec{r})$ is assumed to be very small and completely imaginary. The amplitude of a diffracted electron with respect to the depth within the crystal depends on the amplitudes of all the diffracted electrons. The coupling between the different diffracted beams is determined by the lattice electrostatic potential.
        %
        %
        Starting from \eqref{rele}
        
        \begin{equation}
        \label{kg}
        \Delta \psi + 4 \pi^2 u_0^2 \psi 
        = 
        - \frac{8 \pi^2 m e}{h^2}  V_c (\vec{r})\psi
        .
        \end{equation}
        
        Where 
        
        \[
        V_c (\vec{r}) = V_0 + V'(\vec{r}) + i W (\vec{r}) 
        = V_0 
            + \sum_{\vec{g}\neq 0}
            V_{\vec{g}}
            e^{2 \pi i  \vec{g}\ \vec{r}}
            + i
            \sum_{\vec{g}}
            W_{\vec{g}}
            e^{2 \pi i  \vec{g}\vec{r}}
            =
            \frac{h^2}{2me}
            [Y(\vec{r}) + iY'(\vec{r})]
            ,\]

        \[
        Y(\vec{r}) 
        = 
        \frac{2me}{h^2}
        V_0 + V'(\vec{r})
        =
        \frac{2me}{h^2}V(\vec{r}) 
        = 
        \frac{\sigma}{\pi \lambda} V(\vec{r}).
        \]

        
%
%
%
%
%
%
%

Assuming the primary axis along which the electrons travel to be the $z$-axis, it can be separated from the $x$ and $y$ axes. The wave function after passing through the specimen is the solution of the equation
        
        \[
        \frac{\partial^2 \psi}{\partial z^2} +i 4 \pi u_{0z} +\triangle_{xy} \psi +i4\pi \vec{u}_{xy} . \Delta_{xy} \psi
        =
        -4\pi^2 [Y(\vec{r}) + iY'(\vec{r})] \psi
        \]
        
        Assuming electrons to be of high energy such that back scattering is negligible and the second derivative along the $z$-axis vanishes
        
        \[
        \frac{\partial \psi}{\partial z} = \frac{i}{ 4 \pi u_{0z}} \left[ \triangle_{xy} \psi +i4\pi \vec{u}_{xy} . \Delta_{xy} 
        \right]
        \psi
        +
        4\pi^2 [Y(\vec{r}) + iY'(\vec{r})] \psi
        =
        (\bar{\triangle} + \bar{V})\psi.
        \]
        
        Where, 
        \[ \bar{\triangle}
        =
        \frac{i}{ 4 \pi u_{0z}} \left[ \triangle_{xy} \psi +i4\pi \vec{u}_{xy} . \Delta_{xy} 
        \right]
        , \]
        
         corresponds to the 
         projection 
         of the electron. This is  the propagation operator. In the empty lattice model, its exponential corresponds to the Fresnel propagator which will be introduced in the imaging with a lens section (2.3).  
        The operator 
        
        \[ \bar{V}
        =
        \frac{i\pi}{u_{0z}} [Y(\vec{r}) + iY'(\vec{r})]
        \]
        
        is the electron interaction term. 
        The wave function is given by
        
        \[
        \psi = e^{\int_0^{z_0} \bar{V}(x,y,z) dz} = e^{i \sigma V_p}, 
        \]
        
        where $\sigma = \frac{2 \pi r m e \lambda}{h^2}$ is the interaction constant, and $V_p$ is the projected potential.
        This is what is known as the object transfer function (OTF). The ability to reconstruct the complex $\psi$ uniquely, allows us to determin the potential of the specimen.
        
        It is typical for electron scattering by the specimen to be treated as a multi-slice process where the specimen is considered as layers of potential with thicknesses $\epsilon_i$. The electrons interact with each layer. Disregarding geometrical considerations ($\bar{\Delta} = 0$),the electron propagate freely to the next. The electron wavefunction after passing through a specimen of thickness $z_0$ sliced into $n$ segments is given by \citep{cowleymoodie}
        
        \begin{equation}
        \psi (x,y,z_0)
        =
        e^{\bar{\Delta} \epsilon_n/2}
        e^{i \sigma \bar{V_p}^n}
        e^{\bar{\Delta} \epsilon_{n}/2}
        e^{\bar{\Delta} \epsilon_{n-1}/2}
        e^{i \sigma \bar{V_p}^{n-1}}
        ... 
        e^{\bar{\Delta} \epsilon_1/2}
        e^{i \sigma \bar{V_p}^1}
        e^{\bar{\Delta} \epsilon_1/2}
        \psi(x,y,0).
        \end{equation}
        

        
            Very thin specimens modify only the phase of the electron wave without absorbing any considerable amount. Such objects are known as phase objects. It is reasonable then to consider only the single scattering events of the primary beam. In other words, the electron will not experience multi-scattering within the specimen, what is known as the kinematical approximation
            
            \begin{equation}
            \label{q}
            \psi (\vec{r}) = e^{i \sigma Vp} = q(\vec{r})
			\end{equation}

            For a specimen with low potential, small Z-number,Taylor expansion can  be used to express the electrons' phase, what is known as the \textit{weak phase object approximation}. Taking the amplitude to be unity, the wavefunction is given by;
            
            \begin{equation}
            \label{wpoa}
            \psi (\vec{r}) 
            =
            q(\vec{r})_{WPOA}
            \approx
            1 - \sigma V_p (\vec{r})
            \end{equation}

    \section{The illumination}

The lens is used to form a deamplified image of the electron gun onto the specimen. While the electron gun can be described as a delta function, the illumination on the specimen is given by the convolution between the electron gun and the point spread function of the lens $T(\vec{r})$.
    
    \[
    \psi_i (\vec{r}) 
    = 
    \delta (\vec{r}) 
    \otimes
    T(\vec{r})
    \]
    
    For an ideal lens, $T(\vec{r})$ is the delta function, which is the unitary operator of the convolution. 
    
    For real lenses, $T(\vec{r})$ affects both the phase and amplitude of the electron wave. 
    In momentum space, it is given by 
    
    \[
    T_m (\vec{u}) 
    = 
    e^{i\chi(\vec{u})}
    \]
    
    %
    
    Where the spatial frequency $\vec{u}$ is a 
    vector in the back focal plane of the probe forming lens. $T_m(\vec{u})$ is the \textit{Coherent Transfer Function of the lens}, given by the aberration function $\chi (\vec{u})$. In the case of a perfect lens the transfer function equals one. The  aberrations shift the phase of each of the frequency components of the diffracted electron waves by a different amount.

     It is useful here to introduce some of the main aberrations which must be accounted for in electron microscopy.

     \begin{itemize}
        \item \textbf{Defocus}

    In wave optics, a perfect lens would invert the curvature of a wave created by a point source in order to form a point image. 
    The phase of the spherical wave is constant along spherical surfaces. Therefore it can not be constant along the lens plane.
   The Fresnel propagator describes the phase distribution in the lens plane. In the quadratic approximation of the paraxial behavior, the Fresnel propagator is given by:
    
    \[
    \mathcal{P}_l(x,y) 
    =
    \frac{i}{\lambda l}
    exp 
    \left(
    \pi i \vec{u} \frac{x^2 + y^2}{l}
    \right)    \]
    
    Therefore, using Huygens' principle, the spherical wave front at the plane $z=l$ as a convolution of the wavefront at $z=0$ and Fresnel propagator
    
    \[
    \psi(x,y,l)
    =
    \psi(x,y,0)
    \otimes
    \mathcal{P}_u(x,y) 
    \]

    As a function of the frequency, $\mathcal{P}(\vec{u})_{\triangle f}$ can be expressed as
    \[
    \mathcal{P}(\vec{u})_{\triangle f}
    =
    e^{- i \pi \triangle f u^2} .   
    \]
    
    Where $\triangle f$ is the change in the position of the image plane along the optical axis, known as the defocus.\footnote{Negative $\triangle f$ is called defocus, while positive $\triangle f$ is called overfocus.} 
	%
	%
    %
       Therefore, the phase shift of the  wave 
       due to defocus is given by
    
     \[
     \triangle \chi_{\triangle f}(\vec{u})
    =
    -  \pi \triangle f u^2 .   
    \]

    For a perfect lens, according to the Gaussian imaging, a point in the object plane ($\vec{p}_o$) is conjugate to a point in the image plane ($\vec{p}_i$). The corresponding wave front is spherical. The distance between points in the object is conserved in the image up to a rotation, inversion and magnification. 
    
    However, magnetic lenses are imperfect. For imperfect lenses, the wave front leaving the lens deviates from the spherical wave front. The magnitude of the deviation depends on the  coordinates of the points $\vec{p}_o$, $\vec{p}_i$ and the position of the electron upon passing the lens plane ($\vec{p}_a$).
    
    The result of this is that electrons leaving a point ($\vec{p}_o$) in the object plane within an angular range of a trajectory will form a disk in the image plane. 
    The \textit{aberrated image} of $\vec{p}_o$ is the Union of all disks in the image plane corresponding to $\vec{p}_o$ 

    \begin{figure}[ht]
    \centering
    \includegraphics[scale=2]{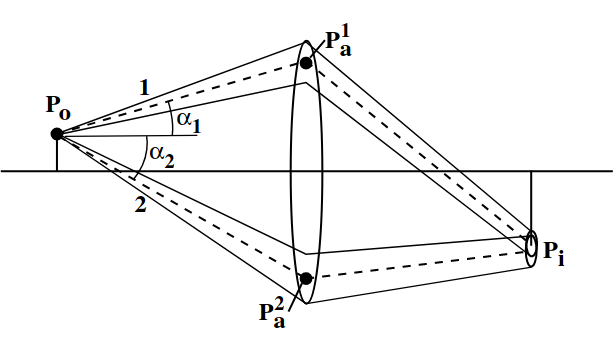}
    \caption{Using an imperfect lens a point object is imaged as a disc. It is called an aberrated image \citep{ictem}. }
    \end{figure}

    Lens aberrations are divided into three categories
    
    \begin{enumerate}
        \item Aperture (geometric) aberrations: 
        are caused by the properties of the magnetic field. They are independent of the object position.

        \item Chromatic aberrations: 
        are caused by the instabilities in the lens current and/or the acceleration potential. They are independent of the position at which the electron passes the lens plane.
        
        \item Parasitic aberrations: 
        result from  the inhomogeneities or imperfections in the lens pole pieces, deviations from perfect circular symmetry, etc. They are linear in either the object position or the position at which the electron passes the lens plane.
    \end{enumerate}{}

    \item \textbf{Primary Seidel Aberrations}

        Since the Gaussian distribution is quadratic, the difference between the wave front exiting the lens and the Gaussian wave front is of third order in its lowest expansion. Primary Seidel aberrations are listed hereafter. 
        They are the 
        most important appercations whose effect needs to be corrected for. Aberrations of higher orders are usually present but their effect is much weaker.
    
        \begin{enumerate}
            \item Spherical aberration,
            
            has coefficient $C_s$ and causes deviation, in the polar coordinates; 

             \[
             \triangle \chi_{C_s}(\vec{u})
            =
            \frac{\pi}{2} C_s \lambda^3 u^4 .   
             \]
            
            \item Coma,
            
            has a coefficient $K$ 
            
             \[
             \triangle \chi_{K}(\vec{u})
            =
            - 2\pi K \lambda^2 \vec{u}^3  cos(\phi - \phi_K).
             \] 
            
            \item Astigmatism and Field curvature,
            
            has coefficients $C_a$ 
            and causes deviation, in the polar coordinates 

             \[
             \triangle \chi_{C_a}(\vec{u})
            =
            - \pi C_a \lambda \vec{u}^2  cos[2(\phi - \phi_a)] .      
             \]
             
             Note that Astigmatism can have higher folds. For example, three fold astigmatism yield a phase shift of 
             \[
             \triangle \chi_{G}(\vec{u})
            =
            - \frac{2\pi}{3} G \lambda^2 \vec{u}^3  cos[3(\phi - \phi_G)].
             \] 
             
            Where, $G$ is the magnitude and $\phi_G$  is the phase angle of the three-fold astigmatism.
             
            \item Distortion,
            
            has coefficients $D,d$. 
            Distortion is only important in low magnification.
            
        \end{enumerate}
        
%
%
    
    
    The total phase shift of the electron wave as a function of the spatial frequency due to aberrations is then: 
    
    \begin{equation}
    \label{chieq}
             \triangle \chi(q)
        =
        - \pi \lambda  q^2 
        \left[ 
        \triangle f 
        + C_a   cos[2(\phi - \phi_a)] 
        \right]
        -\frac{2}{3} \pi \lambda^2  q^3 
        \left[
        G cos[3(\phi - \phi_G)]
        +
        K cos[3(\phi - \phi_K)]
        \right]
        -
        2 \pi C_s
        . 
    \end{equation}


   \item \textbf{Aberration Correction}
    
    For visible light microscopes, aberrations are compensated for by combining lenses of different focal lengths both positive and negative. For electron microscopes some of the aberrations can be corrected physically as follows
    
    \begin{itemize}
        \item Coma can be removed by tilting the incident beam, such that it is aligned along what is known as coma-free axis.
        
        \item Two-fold astigmatism can be corrected using objective stigmator coils. For the three-fold astigmatism, it can be corrected once and for all using special correction coils.
        
    \end{itemize}

        \item \textbf{Scherzer defocus}
    The Scherzer Theorem states that the spherical aberration of a round lens is always a positive quantity. Therefore, spherical aberration causes the rays affected by it to converge before the image plane.
    
    The Scherzer defocus defines the condition for an aperture $H(\vec{u})$ which provides a stable contrast of the image. The Scherzer defocus is given by
    \[
    \triangle f_0 
    = 
    -
    \left(
    1.5 C_s \lambda
    \right)^{1/2}
    \]
    
         \end{itemize}

    \begin{figure}[H]
    \centering
        \includegraphics[scale=0.4]{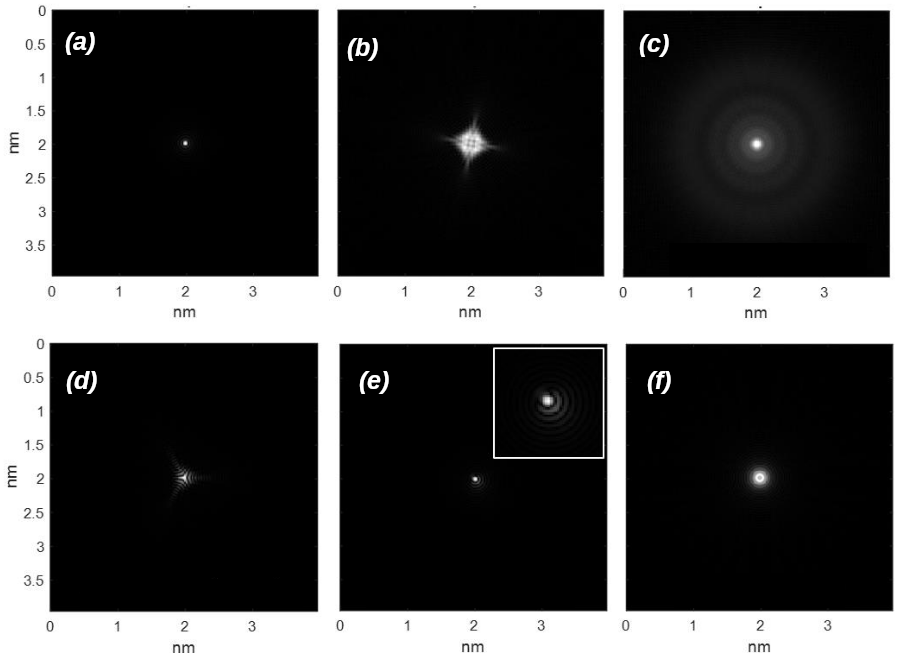}
        \caption{The effect of different aberrations on a free electron beam (a) is visualised in the subsequent images.
        (b) 2-fold astigmatism, 
        (c) spherical aberration, 
        (d) 3-fold  astigmatism, 
        (e) coma, 
        (f) defocus \citep{ltem}.}
    \end{figure}


    \section{The mathematical treatment of Convolution}
    
        The mathematical description of the imaging process is mainly given by convolution of functions in real space where the image exists 
         and their Fourier transform into the momentum space where the diffraction pattern exists. It worth mentioning that the momentum space is also known in the language of microscopy as the spatial frequency space. In this section, the mathematical treatment of convolution is given in details.

        The convolution product, also called folding or flattering, of two functions $f(\vec{r})$ and $g(\vec{r})$ is given by
        \[
        c(\vec{r}) = f(\vec{r}) \otimes g(\vec{r}) = \int \int \int f(\vec{R}) g(\vec{r} - \vec{R}) d \vec{R}
        \]
        
        The identity operator of the convolution is the delta function.
        
        \[
        f(\vec{r}) \otimes \delta(\vec{r}) = \int \int \int f(\vec{R}) \delta(\vec{r} - \vec{R}) d \vec{R} = f(\vec{r})
        \]
        
        This is called the sifting property of the delta function. 
        
        Convolution has two important properties. The first comes from the  multiplication theorem. 
        
        \[
        \mathcal{F}
        \left[
        f(\vec{r})  g(\vec{r})
        \right]
        =
        F(\vec{k}) \otimes G(\vec{k}).
        \]
        
        While the other emerges form the convolution theorem;
        
        \[
        \mathcal{F}
        \left[
        f(\vec{r}) \otimes g(\vec{r})
        \right]
        =
        F(\vec{k})  G(\vec{k}).
        \]
        
        While $\mathcal{F}$ is the Fourier Transform.

\section{The diffraction pattern at the STEM detector}

    The probe is raster scanned across the specimen. At each point, it has position $\vec{R}$ with respect to the specimen. Upon exiting the specimen, the electron wave is given by the multiplication of the illumination function $a (\vec{r}-\vec{R})$ and the object transfere function $q(\vec{r})$ defined in \ref{q}
    
   \[ \psi(\vec{r}) = a (\vec{r}-\vec{R}).q(\vec{r}).\]

    At the detector, the differation pattern is given by the function $M(\vec{u}, \vec{R})$ which is the Fourier transform of the exit wave.
    \[
    M(\vec{u}) 
    = \mathcal{F}[\psi (\vec{r})] 
    = \int \psi (\vec{r}) e^{-2\pi i \vec{u} \vec{r}} d\vec{r}.
    \]

\begin{figure}[H]
    \centering
    \includegraphics[scale=0.45,trim= 3.2cm 0 0 0,clip]{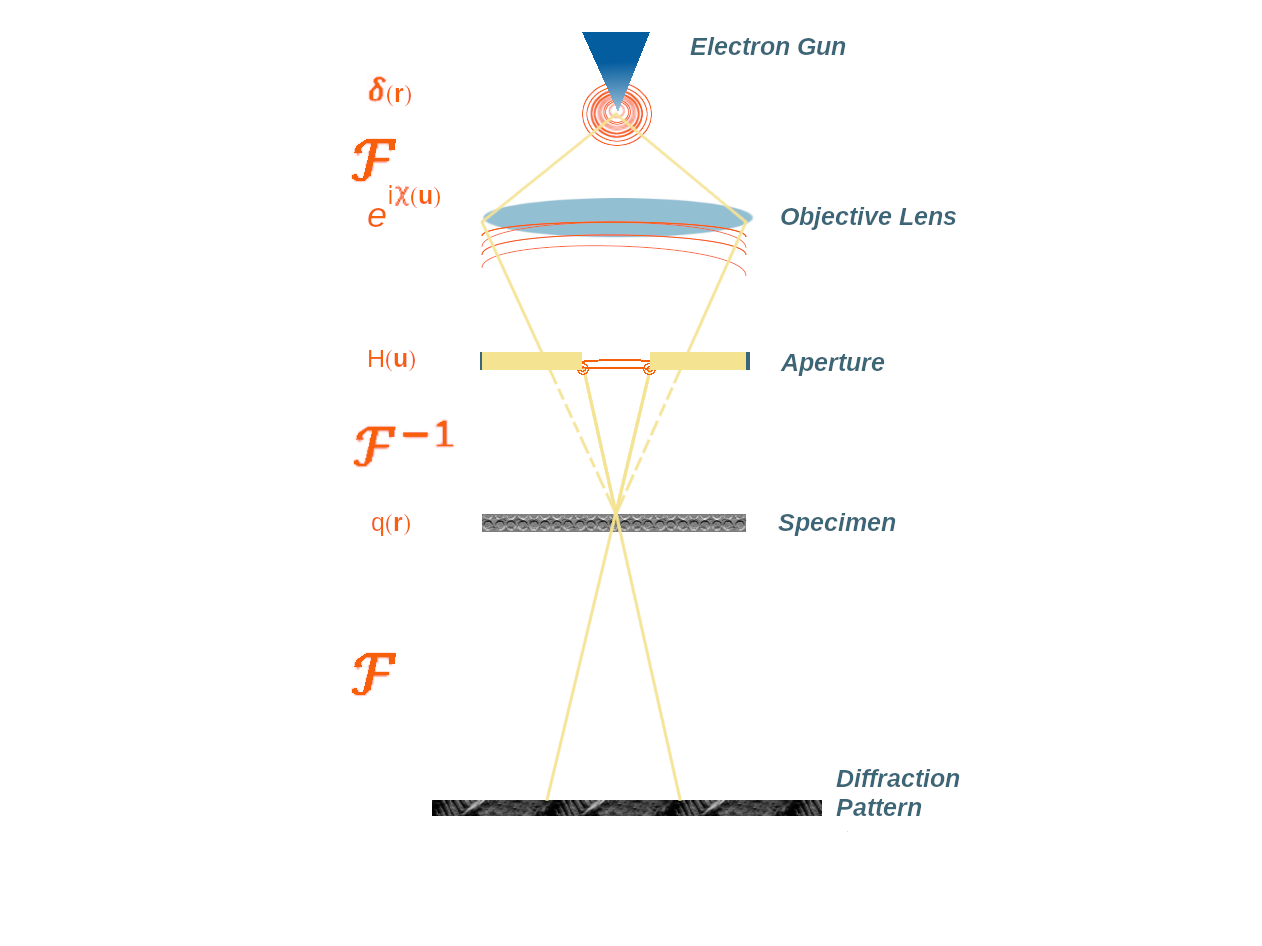}
    \caption{Schematic presentation of STEM imaging. The orange comments describe the evolution of the wave function.}
    \label{stemmath}
    \end{figure}

    A schematic description of the STEM imaging  
    is given in figure \ref{stemmath}. The functions of the individual components and required transformations between the different spaces is annonated.
    
%% file: chapter3.tex


\section{Retrieving the phase and amplitude of the OTF using ptychography}

    \subsection{Ptychographical Iterative Engine (PIE)}
    
    The exit wave $\psi_e$ is the illumination $a$ multiplied by a small part of the probe function $q$.  $\psi_e$ is evaluated with the iteration algorithm with both forward and backward propagation, using the subscript 'NEW' for $q$ and $\psi$ in the latter. Using Wiener filter to avoid the infinity arising from very small probe values, and scaling by the normalized probe modulus we obtain
    
    \begin{equation}
    \label{pie}
     q_{NEW} 
     =
     q 
     +
     \frac{|a|}{|a_{max}|} \frac{a^*}{[|a|^2 + \epsilon]} (\psi_{NEW} - \psi_e )
    =
    q + w \triangle q
    \end{equation}
    
    where $|a_{max}|$ is  the maximum modulus of the probe function and $w$ is the weighing factor.
    
    This update is continuously applied to the same object function for all the probe positions. The whole process is then repeated for typically 50 times, while always refining the estimation of the object.
    
        Several algorithms stem from the ptychograghical iterative engine. For example the extended PIE (ePIE) algorithm uses the normalized probe intensity. A regularized version of the ePIE update is known as rPIE. These algorithms use a tuning parameter as the weighing factor to the probe intensity. 
    \[
    w_{ePIE}
    = \frac{|a|^2}{|a_{max}|^2}
    \]
    
    \[
    w_{rPIE}
    = \frac{|a|^2}{\alpha |a_{max}|^2 + (1-\alpha) |a|^2}
    \]

\begin{figure}[H]
\centering
\includegraphics[scale=.53]{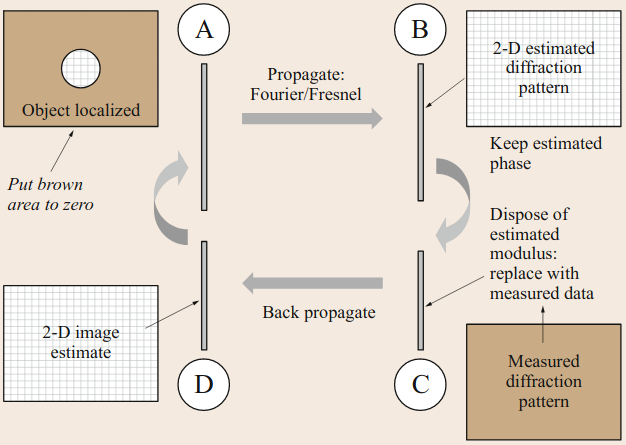}
\includegraphics[scale=.47]{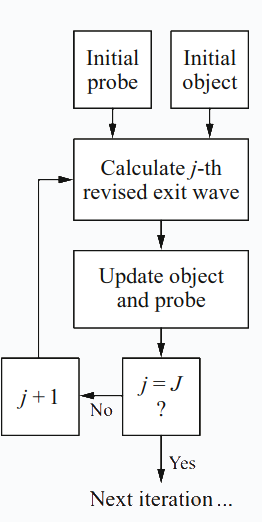}
\caption{For iterative ptychographic algorithms (left) compines the known parameters i.e. the aperture and the differaction pattern and their calculated counter parts through forward and inverse Fourier transform or Fresnel propagator up to a certain number of iterations. While some algorithms take a group of diffraction patterns/ probe positions at the time, ePIE (right) is among the algorithms which treat the differaction patterns sequentially.  }
\end{figure}

\subsection{The Wigner Distribution Deconvolution Algorithm}

Wigner Distribution Deconvolution (WDD) Ptychography has been used for image reconstruction successfully, for example \cite{Yang2016}\cite{Li2014}. It has shown high resolution under high defocus and low dose conditions \cite{dose} 
providing promising potential for biological imaging. 

In this section a detailed explanation, which is mainly mathematical, of the algorithm is given. Since different variables exist in real and momentum spaces it is important to state them clearly. Some of these variables have clear physical meanings, while others are Fourier transforms of physical variables giving raise  to mathematical objects. 
The tables below characterize the functions used in WDD ptychography

\begin{table}[h!]
\center
\caption{variables of real and spatial frequency spaces}
\begin{tabular}{|c|l|c|l|}
	\hline
	symbol & space & dimensionality & definition\\
    \hline 
    $\vec{u}$ & spatial frequency coordinate & 2D & diffraction pattern
    \\ 
    $\vec{R}$ & real space coordinate & 2D & probe position
    \\
    $\vec{r}$ & real space coordinate & 2D & Fourier transform of the diffraction pattern
    \\
    $\vec{U}$ & spatial frequency coordinate & 2D & Fourier transform of probe position
    \\
    \hline
\end{tabular}
\end{table}

\begin{table}[h!]
\centering
\caption{Functions of the WDD algorithm}
\begin{tabular}{|c|c|l|}
\hline
fn. & var. & definition\\
\hline
    $I$ & $(\vec{u},\vec{R})$ & diffraction pattern intensity
    \\ 
    $G$ & $(\vec{u},\vec{U})$ & FT of diffraction pattern intensity w.r.t probe position
    \\
    $H$ & r$(\vec{r},\vec{U})$ & IFT of $G$ w.r.t diffraction pattern
    \\
    \hline
    \end{tabular}
\end{table}
    
    $H$ is a separable function, whose parts are Wigner distribution functions (ambiguity functions), one of the IFT of the diffraction pattern intensity, while the other is FT of the probe position.

    
    The electon wave after exiting the specimen in a STEM column is discribed, as considered already in \ref{q}, by
    \[\psi(\vec{r}) = a (\vec{r}-\vec{R}).q(\vec{r}),\]
    
    where $a$ is the illumination and $q$ is the probe function. In momentum space they are represented by $A$ and $Q$ respectively.
    
    At the detector, the diffraction pattern is 
    
    \[
    M(\vec{u}) 
    = \mathcal{F}[\psi (\vec{r})] 
    = \int \psi (\vec{r}) e^{-2\pi i \vec{u} \vec{r}} d\vec{r}.
    \]
    
    Then the intensity of the diffraction pattern is;
    
    \begin{align*}
    I (\vec{u},\vec{R}) &
    =
    |M(\vec{u},\vec{R})|^2
    =
    |\mathcal{F}[\psi (\vec{r})]|^2
    =
    |A(u) e^{-2\pi i \vec{R} \vec{u}} \otimes Q(\vec{u})|^2
    \\
    &
    =
    \int \int 
    A(\vec{u_a})  Q(\vec{u}-\vec{u_a}) 
    A^*(\vec{u_b})  Q^*(\vec{u}-\vec{u_b})
    e^{-2\pi i \vec{R}(\vec{u_a} - \vec{u_b} ) }
    d\vec{u_a}
    d\vec{u_b}
    \\
    G (\vec{u},\vec{U}) &
    =
    \int I (\vec{u},\vec{R}) e^{-2\pi i \vec{R} \vec{U}} dR
    \\
    &
    =
    \int \int \int
    A(\vec{u_a})  Q(\vec{u}-\vec{u_a}) 
    A^*(\vec{u_b})  Q^*(\vec{u}-\vec{u_b})
    e^{-2\pi i \vec{R}(\vec{u_a} - \vec{u_b} + \vec{U}) }
    d\vec{u_a}
    d\vec{u_b}
    d\vec{R}
    \end{align*}
    
    \begin{align*}
    &
    =
    \int \int 
    A(\vec{u_a})  Q(\vec{u}-\vec{u_a}) 
    A^*(\vec{u_b})  Q^*(\vec{u}-\vec{u_b})
    \delta(\vec{u_a} - \vec{u_b} + \vec{U}) 
    d\vec{u_a}
    d\vec{u_b}
    \\
    &
    =
    \int 
    A(\vec{u_a})  Q(\vec{u}-\vec{u_a}) 
    A^*(\vec{u_a}+\vec{U})  Q^*(\vec{u} - \vec{u_a} - \vec{U}) 
    d\vec{u_a}
    \end{align*}

   \[ \Rightarrow \qquad
    G(\vec{u},\vec{U})
    =
    [A(\vec{u})  
    A^*(\vec{u}+\vec{U}
    ]\otimes_u[
    Q(\vec{u}) 
    Q^*(\vec{u}-\vec{U})
    ]
    \]

    For Wigner Distribution Deconvolution, the function $H$ is obtained by deconvoluting $G(\vec{u},\vec{U})$
    
    \begin{align*}
   H(\vec{r},\vec{U})
   =
  \left[
  \int
   Q(\vec{u})      Q^*(\vec{u}-\vec{U})
   e^{2 \pi i \vec{r} \vec{u}}  d\vec{u}
   \right]
   \times
   \left[
   \int
    A(\vec{u})  
    A^*(\vec{U}+\vec{u})
   e^{2 \pi i \vec{r} \vec{u}}  d\vec{u}
    \right]
    \end{align*}
    
    This is a separable function of the OTF and the illumination. Each of which is given by a Wigner distribution;
    
    \[
    \chi_F (\vec{r},\vec{U}) 
    =
   \int
    F(\vec{u})  
    F^*(\vec{u}-\vec{U})
   e^{2 \pi i \vec{r} \vec{u}}  d\vec{u}.
   \]
   
   Where $F$ is either the OTF or the illumination. Then using the Wiener filter to avoid the instability at $\chi_A \approx 0$, the object Wigner distribution is given by:
   
    \[
    \chi_Q (\vec{r},\vec{U}) 
    = 
    \frac{ 
    \chi_A^* H (\vec{r},\vec{U})
    }{
   |\chi_A (\vec{r},\vec{U}) |^2 + \epsilon.
   }
   \]
   
   Where $\epsilon$ is a function of $\vec{r}$ and $\vec{U}$ but usually a small constant is sufficient\cite{tsr}.
   
   If no aberration exists, the OTF can be obtained from transforming $\chi_Q$ again to the momentum space $u$. The resulting matrix exists in the same space as $G$. OTF is obtained when $u = 0$.\cite{Yang2016}\cite{Li2014}
   
   \[
   D(\vec{u},\vec{U})
   =
   \mathcal{F} \chi_Q (\vec{r},\vec{U})
   =
   Q(\vec{u}) Q^* (\vec{u}-\vec{U})
   \]
   
   This is known as the D-set, which exists in the same coordinate system as $G$ but with the aperture removed. It describes the phase difference between each pixel and every other pixel in the diffraction pattern.
   
   If the source emits partially coherent radiation;
   
    \[
   D(\vec{u},\vec{U})
   =
   \mathcal{F}[s(\vec{r})]   Q(\vec{u}) Q^* (\vec{u}-\vec{U}).
   \]
   
   Where, $s(\vec{r})$ is the intensity distribution of the source.

    With this method one is able to retrieve both the probe function and the object function.

    \subsection{Single Side-band Ptychography}
    
    If the specimen is a weak phase object , then the only important values are $\vec{u} - \vec{U} =0$ and $\vec{U}=0$. Then $Q(\vec{u})  Q^*(\vec{u}-\vec{U})$ has significant values. $Q$ is negligible elsewhere.
    %
    In this case, a shorter version of the WDD ptychographyical method which employs the phase object approximation can the be used to retrieve the object function. This version is what is known as the single side-band method (SSB).
    
    Under the weak phase approximation, the OTF, as given in  \ref{wpoa}, is $q(\vec{r}) = e^{i \sigma V\vec{r})} \approx 1 + i \sigma V(\vec{r}) := 1 + f(\vec{r})$

Where $V\vec{r})$ is the lattice potential and $\sigma$ is the interaction constant. Then

\begin{align*}
    \hspace*{-0.5cm}
    M (\vec{u},\vec{R})
    &
    = 
    A(\vec{u}) e^{-2\pi i \vec{R} \vec{u}} 
    \otimes 
    \mathcal{F}[1 + f (\vec{r})]
    = 
    A(\vec{u}) e^{-2\pi i \vec{R} \vec{u}} 
    +
    A(\vec{u}) e^{-2\pi i \vec{R} \vec{u}} 
    \otimes 
     F(\vec{u})
     \\
     I (\vec{u},\vec{R}) 
     &
     =
     |M(\vec{u},\vec{R})|^2
     =
     |A(\vec{u})|^2 
     + 
     A(\vec{u}) e^{-2\pi i \vec{R} \vec{u}} 
     [A^*(u) e^{2\pi i \vec{R} \vec{u}}  \otimes_{\vec{u}} F^*(\vec{u})]
     + 
     A^*(\vec{u}) e^{2\pi i \vec{R} \vec{u}} 
     [A(\vec{u}) e^{-2\pi i \vec{R} \vec{u}}  \otimes_{\vec{u}} F(\vec{u})]
     \\
     &
     =
     |A(\vec{u})|^2 
     + 
     A(\vec{u}) e^{-2\pi i \vec{R} \vec{u}} 
     \left[ \int
     A^*(\vec{u}-\vec{u}_1) e^{2\pi i \vec{R} (\vec{u}-\vec{u}+1)}  
     F^*(\vec{u})
     d\vec{u}_1
     \right]
     \\
     & 
     + 
     A^*(\vec{u}) e^{2\pi i \vec{R} \vec{u}} 
     \left[ \int
     A(\vec{u}-\vec{u}_1) e^{-2\pi i \vec{R}( \vec{u} -\vec{u}_1)}  
     F(\vec{u})
     d\vec{u}_1
     \right]
     \\
     G (\vec{u},\vec{U}) 
     &
     =
     \mathcal{F}^{-1}[  I (\vec{u},\vec{R})
     ]_{\vec{R}}
     \\
     &
     =
     |A(\vec{u})|^2 
     e^{-2\pi i \vec{R} \vec{U}}
     d\vec{R}
     + 
     \int
     A(\vec{u}) e^{-2\pi i \vec{R} \vec{u}} 
     \left[ \int
     A^*(\vec{u}-\vec{u}_1) e^{2\pi i \vec{R} (\vec{u}-\vec{u}+1)}  
     F^*(\vec{u})
     d\vec{u}_1
     \right]
     e^{-2\pi i \vec{R} \vec{U}}
     d\vec{R}
     \\
     & 
     + 
     \int
     A^*(\vec{u}) e^{2\pi i \vec{R} \vec{u}} 
     \left[ \int
     A(\vec{u}-\vec{u}_1) e^{-2\pi i \vec{R}( \vec{u} -\vec{u}_1)}  
     F(\vec{u})
     d\vec{u}_1
     \right]
     e^{-2\pi i \vec{R} \vec{U}}
     d\vec{R}
     \\
     &
     =
     |A(\vec{u})|^2 \int
     e^{-2\pi i \vec{R} \vec{U}}
     d\vec{R}
     + 
     \int
     A(\vec{u}) e^{-2\pi i \vec{R} \vec{u}} 
     \left[ \int
     A^*(\vec{u}-\vec{u}_1) e^{2\pi i \vec{R} (\vec{u}-\vec{u}_1)}  
     F^*(\vec{u})
     d\vec{u}_1
     \right]
     e^{-2\pi i \vec{R} \vec{U}}
     d\vec{R}
     \\
     & 
     + 
     \int
     A^*(\vec{u}) e^{2\pi i \vec{R} \vec{u}} 
     \left[ \int
     A(\vec{u}-\vec{u}_1) e^{-2\pi i \vec{R}( \vec{u} -\vec{u}_1)}  
     F(\vec{u})
     d\vec{u}_1
     \right]
     e^{-2\pi i \vec{R} \vec{U}}
     d\vec{R}
     \\
     &
     =
     |A(\vec{u})|^2 \delta(\vec{U})
     + 
     \int
     A(\vec{u}) 
     \left[ \int
     A^*(\vec{u}-\vec{u}_1) e^{-2\pi i \vec{R} (\vec{U}-\vec{u}_1)}  
     F^*(\vec{u})
     d\vec{u}_1
     \right]
     d\vec{R}
     \\
     & 
     + 
     \int
     A^*(\vec{u}) 
     \left[ \int
     A(\vec{u}-\vec{u}_1) e^{-2\pi i \vec{R}( \vec{U} -\vec{u}_1)}  
     F(\vec{u})
     d\vec{u}_1
     \right]
     d\vec{R}
     \\
     &
     =
     |A(\vec{u})|^2 \delta(\vec{U})
     + 
     A(\vec{u}) 
     \left[ \int
     A^*(\vec{u}-\vec{u}_1) \delta (\vec{U}-\vec{u}_1)
     F^*(\vec{u})
     d\vec{u}_1
     \right]
     + 
     A^*(\vec{u}) 
     \left[ \int
     A(\vec{u}-\vec{u}_1) \delta( \vec{U} -\vec{u}_1) 
     F(\vec{u})
     d\vec{u}_1
     \right]
     \\
     &
     =
     |A(\vec{u})|^2 \delta(\vec{U})
     + 
     A(\vec{u})
     [A^*(\vec{u} + \vec{U}  )
     F^*(\vec{-U})]
     + 
     A^*(\vec{u})
     [A(\vec{u} - \vec{U}  )
     F(\vec{U})]
\end{align*}
   
   Therefore, the object function can be retrieved by integrating the douple overlap between discs in momentum space of both the the diffraction pattern and the Fourier transform of the probe position $G(\vec{u},\vec{U})$ at $\vec{U} = 0$ and those shifted by a value $\pm \vec{U}$. Nevertheless, the probe function in this case is not retrivable.
   
   \subsection{Schematic presentation of the WDD algorithm}
   
   The schematic figure below explains the programming steps to employ the WDD algorithm in a python script. The subscripts in the variables are used to indicate the original variable. 

   \subsection{Advantages of the dense sampling methods}
   Applying WDD allows us to
   \begin{enumerate}
        \item Cope with the aberrations in the optics after acquisition,
        
        \item Handle strongly diffracting specimen but not multible scattering,
        
        \item Exploit the dark-field intensity, which provides higher resolution  than expected from the probe forming optics,
        
        \item Remove the effects of partial coherence,
        
        \item Choose to image specific layers in 3D specimen, i.e. adapt focus after acquisition. 
        
        \item Obtain higher contrast in the resulting image, 
        
        \item get fast reconstruction (in comparison to iterative ptychography), and
        
        \item obtain high dose efficiency.
        
        
   \end{enumerate}

   \subsection{Promising features}
   
    \begin{itemize}
        \item 3D imaging \cite{epm3d17}
        \item State mixtures imaging
        \item Very simple table-top ptychography microscopes
    \end{itemize}

   \subsection{Notes}
   
   \begin{itemize}
   \item Even though ptychography returns the probe function to a high accuracy, some ambiguities are still irremovable 
   via ptychography.
   
   \item It has not been possible to mathematically prove the correctness of ptychography.
   \end{itemize}
   
   \cite{pty2019}

   \begin{figure}[H]
        \centering
        \includegraphics[scale=1]{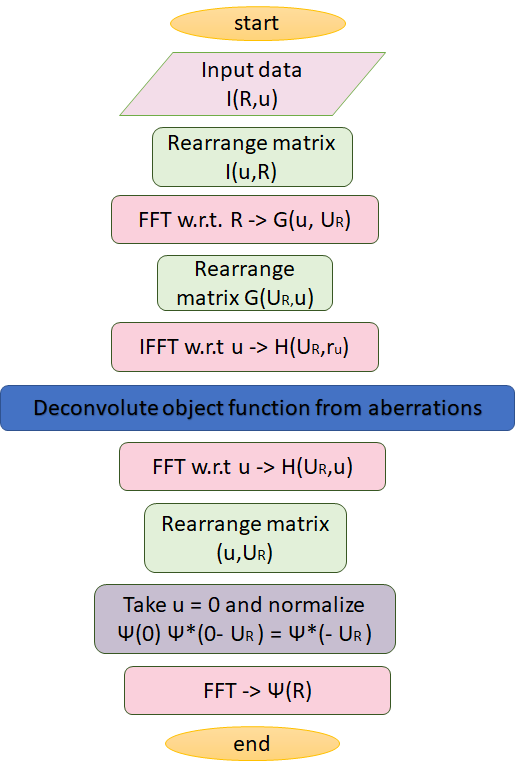}
        \caption{Steps of the WDD ptychographic reconstruction algorthim as it maybe implemented in python}
    \end{figure}

   \begin{figure}[H]
        \centering
        \includegraphics[scale=1]{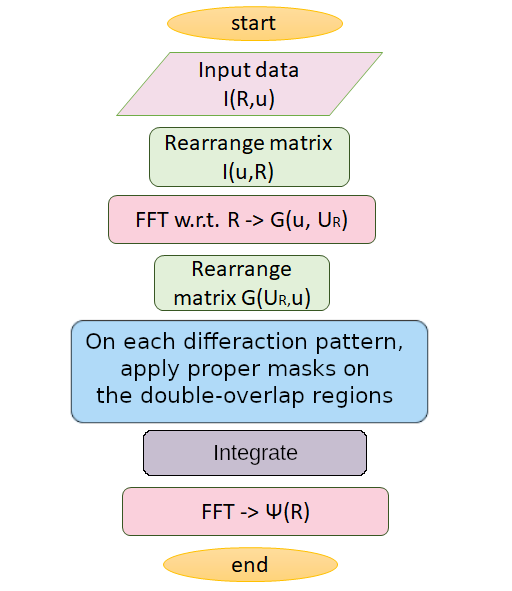}
        \caption{Single Side-Band: Steps of the simpler version of WDD ptychographic reconstruction algorthim *SSB algorithm) as it maybe implemented in python}
    \end{figure}

   

%% file: results.tex
\chapter{Results}

I was provided a 4D simulated data of MoS2 specimen under several conditions. The goal was to develop a python program and to use it to produce the phase images of these data sets. Due to the new condition of the diploma thesis as only literature review and a 2-weeks time constraint for the partial work in the host group in J\"{u}lich that has been set on the practical phase, this goal was only partially achieved. In this chapter the images produced using the python code I wrote are presented.  In the second section images form a common laboratory session I participated in are presented. They are compared to the images produced using a program developed by the moreSTEM group which employs the ePIE method.

\section{WDD reconstruction of MoS2}

    Molybdenum disulfide (MoS2) is considered to be a  promising ultrathin layered semiconductor. It has a direct band gap in the monolayer. 
     It's few-layers regime has promising  applications in nanoelectronics, optoelectronics,  flexible devices as well as spintronic and valleytronic devices \cite{mos2review}.   Its structure  is shown below as given by \cite{mos2struct}
    
   
    
    
     \begin{figure}[H]
     \centering
         \includegraphics[scale=1]{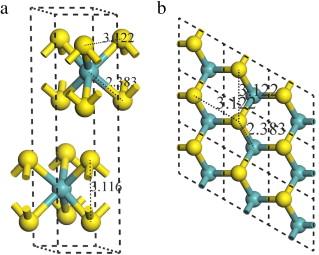}
         \caption{Structure of MoS2 (\cite{mos2struct})}
     \end{figure}

    A 4D data set was simulated using Multislice Frozen Phonon algorithm by \cite{ben}. It represents 43 x 25 scan positions and 236 x 236 detector pixels. The illumination angle is 24 mrad, the illumination step size is 12.644pm, while the electron acceleration is 80keV equivalent to an electron wavelength of $\lambda$ = 4.176pm

    
    The dataset was used to produce conventional STEM images by simulating virtual detectors.
    
    \begin{figure}[H]
    \centering
    \includegraphics[scale=3.6,trim = 0 0 0 0 ,clip]{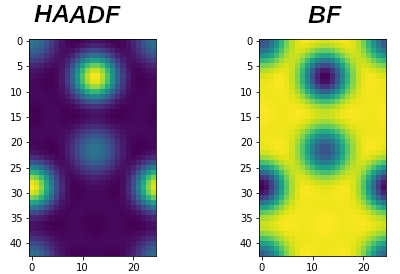}
    \caption{Conventional bright field (BF) and high angle annular darrk field (HAADF) STEM image of MoS2 monolayer are produced from the 4D dataset}
    \label{stemMoS2}
    \end{figure}

    
%
    The single side-band method was used on 
    data set to produce amplitude and phase images of the monolayer MoS2 specimen
    \begin{figure}[H]
    \centering
    \includegraphics[scale=.9,trim = 0 0 0 0 ,clip]{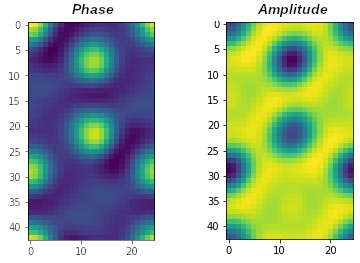}
    \caption{amplitude and phase images of monolayer MoS2 using single side-band ptychography}
    \label{ssbMoS2}
    \end{figure}
    
    The simulated probe is aberration-free. Therefore for the SSB reconstruction, only the aperture is taken into consideration.

\section{Polycristallize Gold Specimen}

    \begin{figure}[H]
    \centering
    \includegraphics[scale=0.55,trim = 0 0 0 0 ,clip]{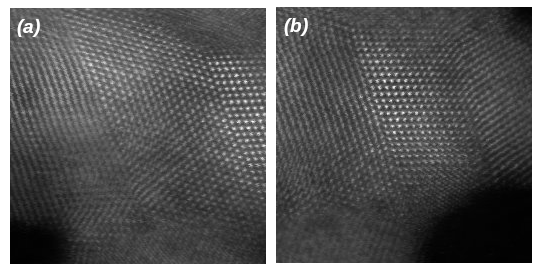}
    \caption{ADF images of characterizing gold grid usually used for the FEI TITAN 30-800 STEM aberration correction. (a) an optimal image with aberrations appropriately corrected. (b) Amplified spherical aberration (C$_s$ = 6 $\mu$m  }
    \label{adfgold}
    \end{figure}
    
     The images \ref{adfgold} was obtained on a laboratory session together with H. L. Robert at The Ernst Ruska-Centre for Microscopy and Spectroscopy with Electrons on July 2020. They are obtained using Annular Dark Field detector.

    These images can be compared to the images produced using ePIE by \cite{achim} shown below
    
    \begin{figure}[H]
    \centering
    \includegraphics[scale=.27,trim = 0 0 0 0 ,clip]{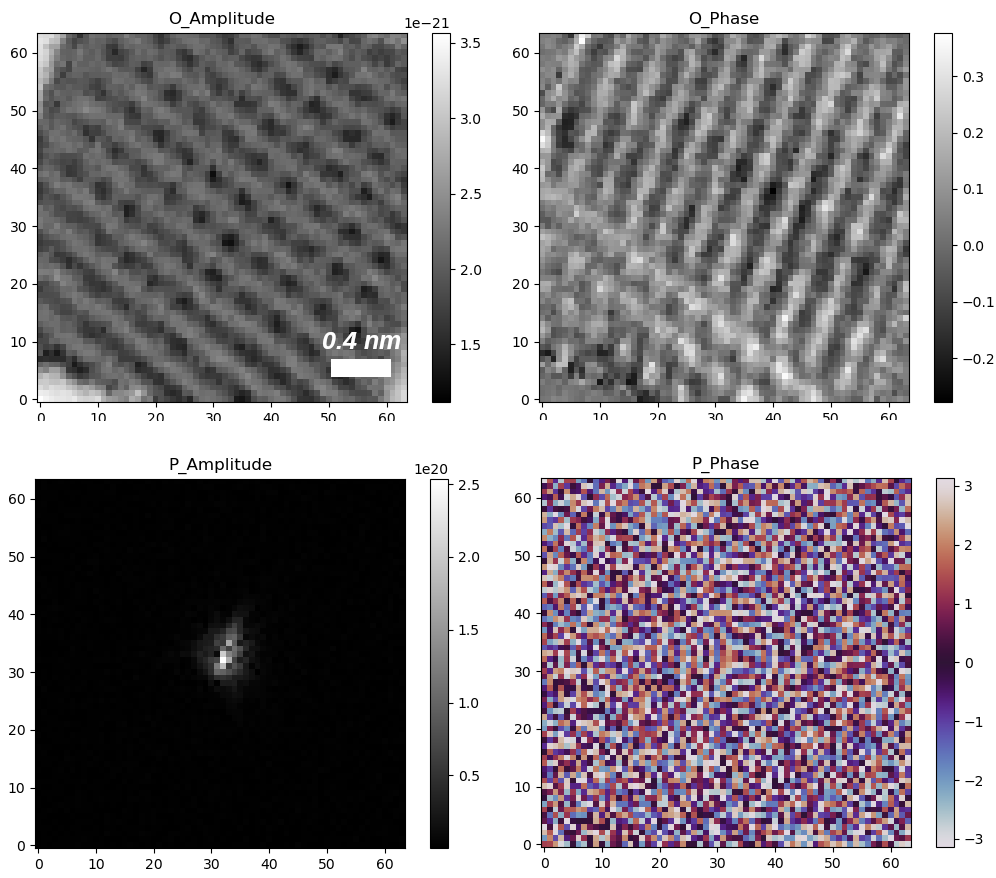}
    \caption{Reconstruction of the aberration-free gold image (a in figure \ref{adfgold}) using ePIE. Ptychography retrieves the phase and amplitude of both the object (O) and the probe (P) \cite{achim}.}
    \label{epie1}
    \end{figure}

    \begin{figure}[H]
    \centering
    \includegraphics[scale=.27,trim = 0 0 0 0 ,clip]{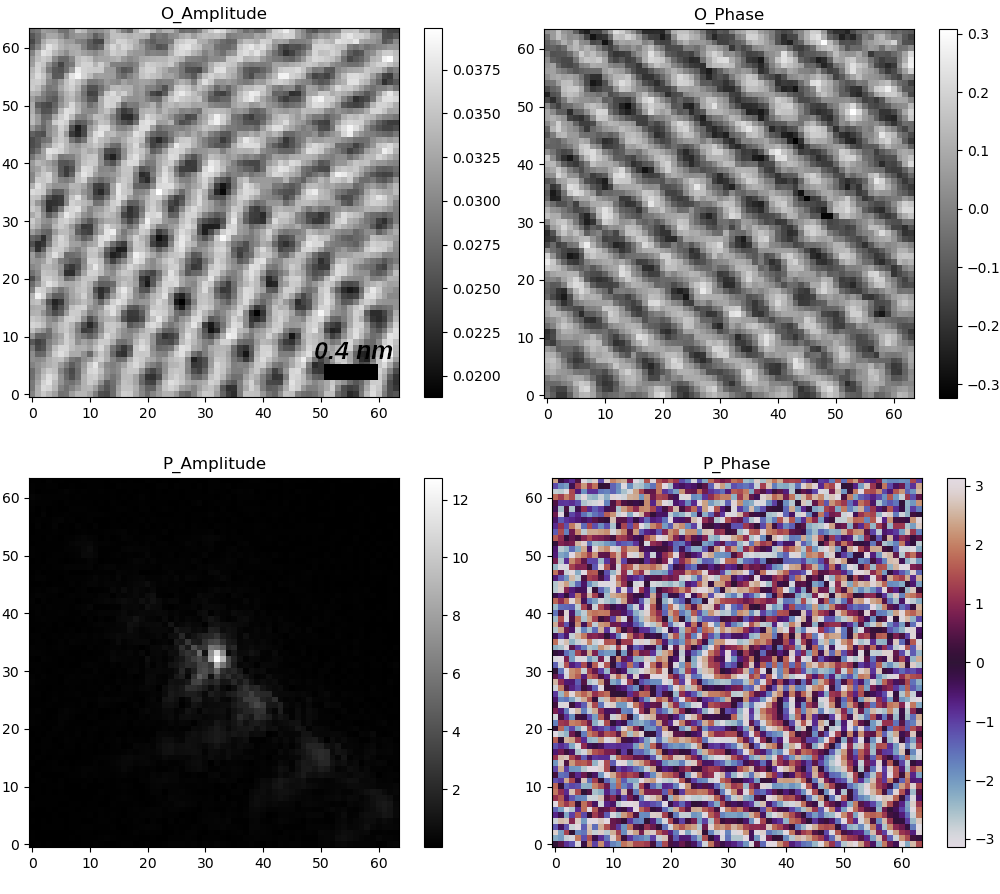}
    \caption{Reconstruction of the aberrated gold  image (b in figure \ref{adfgold}) using ePIE. Ptychography retrieves the phase and amplitude of both the object (O) and the probe (P) \cite{achim}.}
    \label{epie2}
    \end{figure}

\section{Conclution}
   
Ptychography is a powerful tool for microscopy imaging including
STEM. There are several methods of STEM ptychography all capable of
retrieving the complex transfer function of the object, and therefore
its potential. The reconstructed images have super resolution. 
Nevertheless, ptychography is applicable on very thin specimens with
thickness 2-5 nm and low atomic number. It is inapplicable in the
case of multiple scattering. 
The recording speed of Ptychography is still an open question where
improvements are made to reach the goal of live ptychography.